\pdfoutput=1
 \documentclass[prb,twocolumn,showkeys,showpacs,superscriptaddress]{revtex4}
\usepackage{epsfig,latexsym,amssymb}

\begin{document}

\title{Unified Phase Diagram for the Three-Dimensional XY Model of a Point Disordered Type-II Superconductor}

\author{Peter Olsson}
\affiliation{Department of Physics, Ume{\aa} University, 901 87 Ume{\aa}, Sweden}
\author{S. Teitel}
\affiliation{Department of Physics and Astronomy, University of
Rochester, Rochester, NY 14627}
\date{\today}

\begin{abstract}
We carry out extensive Monte Carlo simulations of the three-dimensional (3D) uniformly frustrated XY model with uncorrelated randomly perturbed couplings, as a model for the equilibrium behavior of an extreme type-II superconductor with quenched uncorrelated random point vortex pinning, in the presence of a uniform applied magnetic field.  We map out the resulting phase diagram as a function of temperature $T$ and pinning strength $p$ for a fixed value of the vortex line density.  At low $p$ we find a sharp first order vortex lattice melting phase boundary separating a vortex lattice from  a vortex liquid.  As $p$ increases, it appears that this first order transition smears out over a finite temperature interval, due to the effects of the random pinning, in agreement with several recent experiments.  At large $p$ we find a second order transition from vortex liquid to vortex glass.  
\end{abstract}
\maketitle

\section{Introduction}

In strong type-II superconductors, where the magnetic penetration length $\lambda$ is much larger than the bare coherence length $\xi_0$, much of the macroscopic behavior can be modeled in terms of the interacting vortex lines  that are introduced into the system by the application of an external magnetic field.\cite{Tinkham}  In the high temperature copper-oxide superconductors such as YBCO and BSCCO, material parameters such as high anisotropy, as well as the high transition temperatures, cause thermal fluctuations to play an important role.  Such materials are also believed, even in the purest of single crystal samples, to contain quenched, intrinsic, uncorrelated,  random point impurities that can pin vortex lines.  The resulting phase diagram as a function of temperature and applied magnetic field is generally believed to result from such a combination of thermal fluctuations and random point pinning.\cite{Blatter} It was later argued that similar fluctuation effects can be observed even in single crystal samples of more familiar strong type-II low temperature superconductors\cite{ Paltiel, Ling, Banerjee, Marchevsky, Pasquini} such as Nb and NbSe$_2$, albeit over a much more restricted region of the phase diagram.  It has been argued that a universal phase diagram might apply to all strong type-II superconductors.\cite{Menon}

Considerable experimental effort has been devoted to the determination of this vortex line phase diagram.  \cite{Worthington,Liang, Schilling, Khaykovich, Nishizaki, Bouquet, vanderBeek, Gaifullin, Avraham, Radzyner, LiWen, Beidenkopf, Beidenkopf2, Petrean, Andersson, Strachan, Strachan2, Shibata, Soibel}  
It is now generally accepted that upon increasing temperature at low magnetic fields, a sharp first order melting transition exists\cite{Liang, Schilling, Khaykovich, Nishizaki, Bouquet} from an elastically distorted vortex lattice, known as the ``Bragg glass,"\cite{Giamarchi,Nattermann,FisherBG} to a disordered vortex liquid.
However many other aspects of the vortex phase diagram remain in dispute.  At low temperatures, increasing the magnetic field leads to a first order transition from the Bragg glass to a disordered vortex state.\cite{vanderBeek, Gaifullin} It was originally believed that a special critical point separated the low field thermally induced vortex lattice melting from the high field disorder (random pin) induced melting.\cite{Khaykovich,Nishizaki, Bouquet}  However later work\cite{vanderBeek, Gaifullin, Avraham, Radzyner, LiWen} argued for a single unified first order phase boundary for the vortex lattice, continuing smoothly down to low temperatures.
At low temperatures and high magnetic fields, where the vortex lines are spatially disordered, a ``vortex glass" phase has been proposed.\cite{FFH, Scheidl}   It remains in question whether this vortex glass is a truly distinct thermodynamic state with true superconducting phase coherence,\cite{Petrean,Andersson} separated from the vortex liquid by a continuous second order phase transition, or whether there is just a crossover to a highly viscous vortex liquid upon decreasing temperature.\cite{Strachan,Strachan2}  It has also been proposed that  within the vortex liquid there is a sharp first order transition line that splits off from the melting line at higher magnetic fields, and terminates at a critical end point.  This transition is  claimed to separate regions with lesser vs greater spatial vortex correlations, the more correlated region being called the ``vortex slush".\cite{Worthington,Shibata}  Other works have argued that the disordered vortex state at high magnetic fields continues to exist as a thin sliver of ``multidomain glass" all along the vortex melting line, even at low magnetic fields.\cite{Banerjee,Menon}

Theoretically, the vortex line phase diagram has been studied within effective elastic theories, often using the Lindemann criterion to estimate the location of melting and other transitions.\cite{Ertas,Giamarchi2,Koshelev,Kierfeld,Mikitik,Kierfeld2}  Several of these works\cite{Kierfeld,Mikitik, Kierfeld2} reported the possibility of a critical end point to the vortex lattice melting line at high magnetic fields, as well as a vortex slush phase.  However later analytical calculations, based on the Ginzburg-Landau model with pinning,\cite{LiRosen} and on a vortex line model in a weak impurity background with both elastic and plastic excitations,\cite{Dietel} found only a unified first order melting transition for the Bragg glass, extending to low temperatures with no critical points intervening, and no vortex slush.

The difficulty of performing reliable analytical calculations has thus led to numerous numerical investigations,\cite{Ryu,Wilkin,Otterlo,Reichhardt, NonoHu, OTmelt, OlssonVG, Kawamura0, Kawamura, Lidmar, DasguptaValls1, DasguptaValls2} in which one hopes to map out various aspects of the vortex line phase diagram within a well defined simple model.  In this paper we report on one such investigation, using the three-dimensional (3D) uniformly frustrated XY model with uncorrelated quenched disorder in the couplings, as a model for a strong type-II superconductor with random point pinning.  
A virtue of the XY model is that it describes realistic
vortex line interactions with no restrictions on
vortex line excitations; both overhangs in the field induced
vortex lines, and closed thermally excited vortex loops are included.  The details
of vortex line cores are handled by the short length cutoff of the
numerical grid, and hence issues concerning vortex line cutting
are treated with a minimum of ad hoc assumptions.  Our model holds in the limit of infinite penetration length, $\lambda\to\infty$, and we comment later on the implications of this approximation.

\begin{figure}[tbp]
\epsfxsize=8.6truecm
\epsfbox{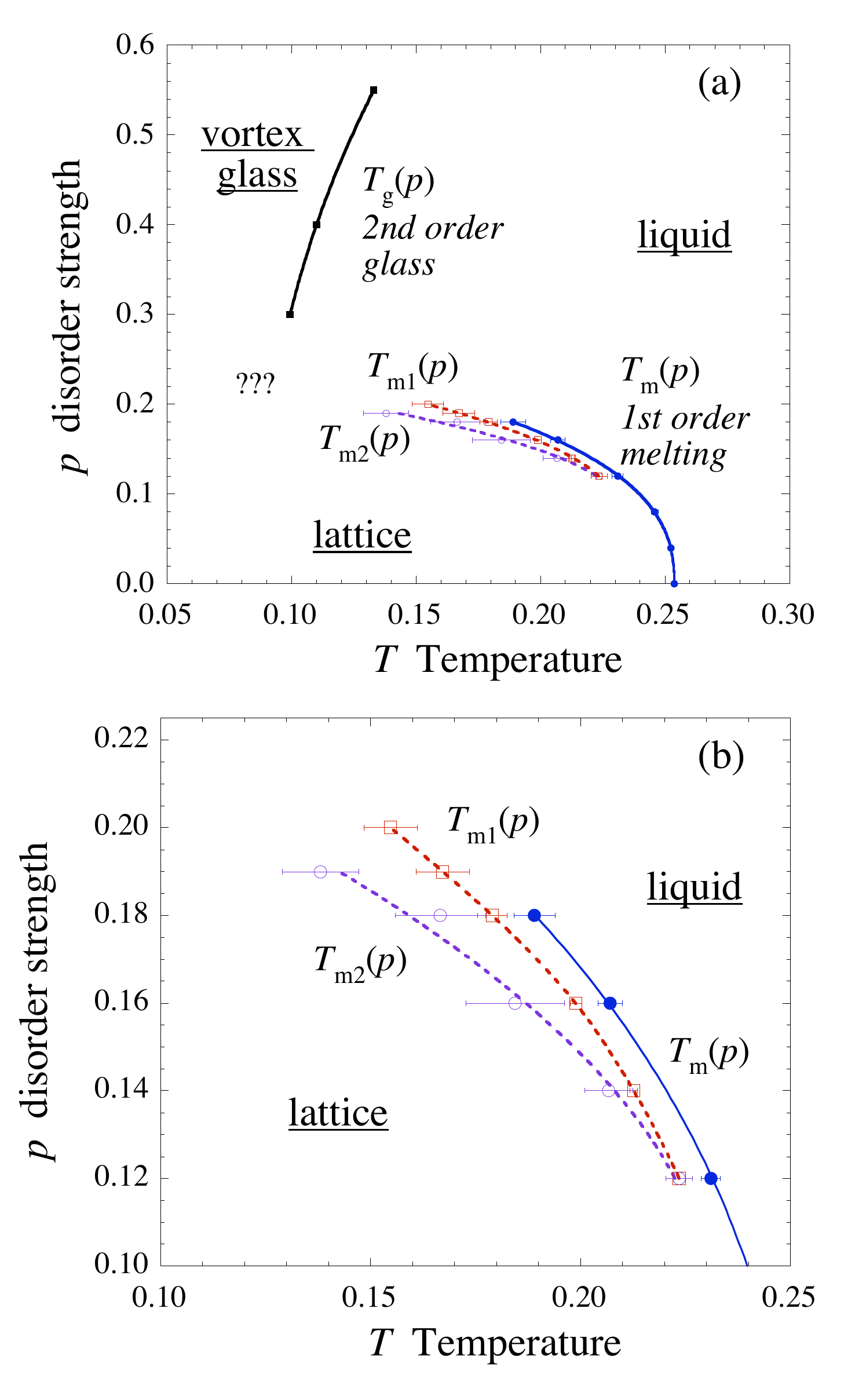}
\caption{(Color online)
a) Phase diagram of the point disordered 3D XY model of Eq.~(\ref{eH}) as a function of temperature $T$ and disorder strength $p$ for fixed vortex line density $f=1/5$ and coupling ratio $J_z=J_\perp/40$. The solid blue circles ($\bullet$) and solid line at low $p$ indicate a sharp first order vortex lattice melting transition $T_{\rm m}(p)$, as determined from a system of size $L=20$, $L_z=6$ and averaged over $8$ independent realizations of the quenched randomness.
As $p$ increases, this melting transition is observed to smear out over a finite temperature interval, $T_{\rm m2}\le T\le T_{\rm m1}$, as determined from a system of size $L=20$, $L_z=12$ and averaged over $4-8$ independent realizations of the quenched randomness; $T_{\rm m1}$ is indicated by the open red squares ($\Box$) and dashed line, while $T_{\rm m2}$ is indicated by the open purple circles ($\circ$) and dashed line. 
Error bars represent the standard deviation of values obtained over the independent realizations.
At large $p$ we find a sharp second order transition $T_{\rm g}(p)$ from vortex liquid to vortex glass, as determined by a detailed finite size scaling analysis of system sizes $L=10-20$, $L_z=(3/5)L$, averaged over $200-600$ independent realizations of the quenched randomness; $T_{\rm g}(p)$ is indicated by the solid black squares ($\blacksquare$) and solid line.  We are unable to equilibrate our system at the low temperatures (denoted by ``???") where the vortex melting and vortex glass transitions approach one another.
b) Expanded view of the phase diagram showing the smearing of the vortex lattice melting transition as $p$ increases.
}
\label{f1}
\end{figure}

We build upon our earlier work\cite{OTmelt, OlssonVG} to present here the equilibrium phase diagram as a function of temperature $T$ and disorder strength $p$, for a fixed density of vortex lines $f=B\xi_0^2/\phi_0=1/5$ ($\phi_0$ is the flux quantum).
Keeping $f$ fixed avoids effects that would be due to varying
commensurability of the vortex lines with respect to the underlying
numerical grid of the model.  However, since increasing $f$ in a continuum system
is generally believed to increase the effective pinning strength, \cite{MenonDasgupta} as more
lines get forced into the same pinning volume, varying $p$ at fixed $f$
should provide qualitatively similar information as the more physical
situation of varying $f$ at fixed $p$.  
Through careful, well equilibrated, simulations comparing systems of different size, and averaging over  different realizations of the random pinning, we consider the limits of both weak and strong disorder.  
Our results are summarized in Fig.~\ref{f1}.
By using a denser vortex line system than in our earlier work we are able in particular to explore the strong pinning limit, correcting our earlier preliminary conclusions\cite{OTmelt} concerning the presence of a vortex glass in this model.  We also find, for the first time, a disorder induced smearing of the vortex lattice melting transition $T_{\rm m}(p)$ at intermediate disorder strengths, that we believe is in good agreement with recent experiments.\cite{Pasquini, Soibel}

The remainder of this paper is organized as follows.  In Sec.~II we define our model and the quantities we use to map out the phase diagram.  In Sec.~III we present our results for vortex lattice melting in the limit of weak pinning strengths.  In Sec.~IV we present our results for the vortex glass at strong pinning strengths.  In Sec.~V we discuss the smearing of the vortex lattice melting transition at intermediate pinning strengths.  In Sec.~VI we summarize our results and discuss their relation to other recent numerical works.

\section{Model}
\label{smodel}

The model we simulate is the uniformly frustrated 3D XY model, 
given by the Hamiltonian,\cite{YHLi}
\begin{equation}
{\cal H}[\theta({\bf r}_i)]=-\sum_{i\mu}J_{i\mu}\cos(\theta({\bf r}_i)-\theta({\bf r}_i+\hat\mu)-A_{i\mu})\enspace,
\label{eH}
\end{equation}
which serves as a discretized approximation to the Ginzburg-Landau free energy functional in the London approximation
of fixed wavefunction amplitude, and captures the kinetic energy of the flowing supercurrents in the system.
Here $\theta({\bf r}_i)$ is the thermally fluctuating phase angle of the superconducting wavefunction on site ${\bf r}_i$
of a cubic $L\times L\times L_z $ grid of sites with bonds in directions $\hat\mu=\hat x,\hat y,\hat z$.  
$J_{i\mu}$ is the superconducting coupling constant on bond $(i\mu)$ of the grid, $A_{i\mu}$ is 
proportional to the magnetic vector potential integrated across the bond, and the argument of the cosine is thus the
gauge invariant phase angle difference across the bond.

The $A_{\rm i\mu}$ are determined such that
the circulation of $A_{i\mu}$
around any plaquette $\alpha$ of the grid is fixed and equal to $2\pi f_\alpha$ with $f_\alpha$ equal to the fraction of applied magnetic flux quanta through that plaquette.  To model a uniform applied magnetic field in the $\hat z$ direction, 
we use a uniform value $f_\alpha = f$  for plaquettes oriented with normal in the $\hat z$ direction, and 
$f_\alpha =0$ otherwise.  The presence of the applied magnetic field induces vortex lines into the phase angles 
$\theta({\bf r}_i)$ such that the average vortex line density in direction $\hat \mu$ is $n_\mu=f\delta_{\mu z}$.
Holding the $A_{i\mu}$ fixed corresponds to the approximation of an infinite magnetic penetration length 
$\lambda$, which one may expect to be reasonable in the limit that $\lambda$ is much larger than the 
average inter-vortex spacing.  
Experiments on very pure YBCO samples\cite{Liang} indicate that many features of behavior,
up to surprisingly large magnetic fields, are well described by this
large $\lambda\to\infty$ XY model limit. 
We return to comment on this approximation in
section \ref{svg}.

To model vortex pinning due to quenched point randomness we use couplings\cite {OTmelt} 
\begin{eqnarray}
  \label{eJ}
   J_{i\mu}&=&J_\perp(1+p\epsilon_{i\mu}),\quad \mu=x,y\\ \nonumber
  J_{iz}&=&J_z  \quad {\rm constant},\end{eqnarray}
where the $\epsilon_{i\mu}$ are uncorrelated, uniformly distributed, random
variables with
\begin{equation}
 \langle\epsilon_{i\mu}\rangle=0, \qquad \langle\epsilon_{i\mu}^2\rangle=1\enspace. 
\end{equation}
The disorder strength is thus controlled by the parameter $p$ in Eq.~(\ref{eJ}).
We will vary $p$ from small to large values in order to systematically investigate
the differences between weak and strong pinning.
 
In the following we will use $\langle\dots\rangle$ to denote the equilibrium average for 
a particular realization of the random pinning $\{\epsilon_{i\mu}\}$.  
We will use $[\dots ]$ to denote the average over several
independent realizations of the $\{\epsilon_{i\mu}\}$. 

In this work we use parameters,
\begin{eqnarray}
f&=&1/5\enspace,\\
J_z&=&J_\perp/40\enspace.
\label{eparms}
\end{eqnarray}
We use $J_z\ll J_\perp$ to enhance vortex line fluctuations along the $\hat z$ direction,
thus allowing us to use systems with smaller $L_z$.  
The relatively dense value of $f$ was similarly chosen so as to have many vortex
lines contained within systems of modest size, so as to permit finite size scaling analyses
(particularly for the vortex glass) and to average over many independent realizations of
the random disorder $\{\epsilon_{i\mu}\}$.  While this high density leads to 
artificial commensurability effects (for example a square rather than a triangular
ground state vortex lattice), one may still hope that many features of our 
system with respect to vortex lattice melting and the vortex glass transition will
remain qualitatively similar to results on real physical materials. 

To probe the behavior of the system we consider the average energy per site,
which is the thermodynamic conjugate variable to the temperature $T$,
\begin{eqnarray}
E&\equiv& {1\over L^2L_z}{\partial (\beta {\cal F})\over \partial \beta}\\ \nonumber
&=&-{1\over L^2L_z}\sum_{i,\mu}J_{i\mu}\langle \cos(\theta({\bf r}_i)-\theta({\bf r}_i+\hat\mu)-A_{i\mu})\rangle \enspace,
\end{eqnarray}
where $\beta\equiv1/T$ and ${\cal F}$ is the total free energy.  We can similarly define a variable $Q$ which is the
thermodynamic conjugate to the disorder strength $p$,
\begin{eqnarray}
Q&\equiv&{1\over L^2L_z}{\partial {\cal F}\over\partial p}\\ \nonumber
&=&-{1\over L^2L_z}\sum_{i,\mu=x,y}J_\perp\epsilon_{i\mu}\langle\cos(\theta({\bf r}_i)-\theta({\bf r}_i+\hat\mu)-A_{i\mu})\rangle \enspace.
\end{eqnarray}

The vorticity in the system is determined by considering the circulation of the gauge invariant phase
difference around each plaquette.  For a plaquette $\alpha$ at position ${\bf r}_\alpha$ with normal in direction $\mu$,
the vorticity $n_\mu({\bf r}_\alpha)$ piercing the plaquette is determined by,
\begin{equation}
\sum_\alpha \left[\theta({\bf r}_i)-\theta({\bf r}_i+\hat\mu)-A_{i\mu}\right]_{-\pi}^{+\pi} = 2\pi[n_\mu({\bf r}_\alpha)-f\delta_{\mu z}]\enspace,
\end{equation}
where the sum goes counterclockwise around the bonds of the plaquette, and
the bracket on the left hand side indicates that the gauge invariant phase angle difference is to be computed so as to lie within the interval $(-\pi,\pi]$. 

To look for vortex lattice ordering we compute the vortex structure function
$S({\bf k}_\perp)$ which measures correlations between vortices within the same $xy$ plane,
\begin{equation}
S({\bf k}_\perp)={1\over fL^2L_z}\sum_{{\bf r}_\perp,{\bf r}_\perp^\prime,z}{\rm e}^{i{\bf k}_\perp\cdot({\bf r}_\perp-{\bf r}_\perp^\prime)}\langle n_z({\bf r}_\perp,z)n_z({\bf r}_\perp^\prime,z)\rangle
\label{eSk}
\end{equation}
where ${\bf r}_\perp\equiv (x,y)$ denotes the coordinates in the $xy$ plane
and similarly ${\bf k}_\perp \equiv (k_x,k_y)$.  The presence of a vortex lattice will be indicated
by the appearance of sharp peaks in $S({\bf k}_\perp)$ at reciprocal lattice vectors ${\bf K}$.

To look for a possible vortex glass phase, in which vortex lines are frozen into a disordered configuration
and so $S({\bf k}_\perp)$ displays no signal of the ordering, we consider the helicity modulus.\cite{OlssonVG}
To simulate the Hamiltonian of Eq.~(\ref{eH}) on a finite size grid, we use fluctuating twist boundary
conditions, defined by,
\begin{equation}
\theta ({\bf r}_i+L_\mu\hat\mu)-\theta({\bf r}_i)=\Delta_\mu\enspace,
\end{equation}
where $\Delta_\mu$, the total phase angle twist across the system in direction $\hat \mu$,
is taken as a thermally fluctuating degree of freedom.  We then compute the histogram
$P(\Delta_\mu)$ of values that $\Delta_\mu$ takes during the course of the simulation 
(averaging over the twists in the transverse directions) and
define the free energy variation with twist by, ${\cal F}(\Delta_\mu)\equiv -T\ln P(\Delta_\mu)+{\rm constant}$.
The helicity modulus $\Upsilon_\mu$ in direction $\hat \mu$ is then defined in terms of the curvature of ${\cal F}(\Delta_\mu)$ at its minimum $\Delta_{\mu 0}$,
\begin{equation}
\Upsilon_\mu\equiv {L_\mu\over L_\nu L_\sigma}\left.{\partial^2{\cal F}\over\partial\Delta_\mu^2}\right|_{\Delta_\mu=\Delta_{\mu 0}}\enspace,
\label{eUps}
\end{equation}
where $\mu,\nu,\sigma$ are a permutation of $x,y,z$.
Note, unlike pure systems, for random systems it is not generally true that $\Delta_{\mu 0}=0$.
When ${\cal F}(\Delta_\mu)$ varies with $\Delta_\mu$, and so the system is sensitive to the boundary conditions,
we have $\Upsilon_\mu>0$ and the system possesses superconducting phase coherence.
When ${\cal F}(\Delta_\mu)$ is flat and independent of $\Delta_\mu$, $\Upsilon_\mu=0$ and superconducting
phase coherence is lost.  The vanishing of $\Upsilon_\mu$ thus is a signature of the
superconducting transition.

At a second order phase transition $T_{\rm c}$, the temperature and system size dependence of 
$\Upsilon_\mu$ is expected to obey a critical scaling law.  Since the applied magnetic field 
singles out a special direction, there is the possibility that this scaling may be {\it anisotropic}.
In such a case, the expected scaling law for $\Upsilon_\mu(T,L,L_z)$ is,
\begin{equation}
{L_\nu L_\sigma\over L_\mu}\Upsilon_\mu (T,L,L_z) = u_\mu(tL^{1/\nu},L_z/L^\zeta)\enspace,
\end{equation}
where $t\equiv T-T_{\rm c}$, $u$ is the scaling function, $\nu$ is the correlation length
critical exponent, and $\zeta$ is the anisotropy critical exponent.   Should the scaling turn
out to be {\it isotropic}, then $\zeta=1$, and for systems with a fixed aspect ratio
$L_z=\gamma L$ the scaling law reduces to,
\begin{equation}
L\Upsilon_\mu(T,L) =  \tilde u_\mu(tL^{1/\nu})\enspace.
\label{eUpsScale}
\end{equation}
In this case, exactly at the transition temperature $T_{\rm c}$, one has $t=0$ and so 
$L\Upsilon_\mu$ is independent of system size $L$.

Henceforth, we will measure
temperature and energies in units where $J_\perp=1$.  Length will be measured in units where the grid spacing is unity.

\section{Vortex Lattice Melting at Weak Pinning}
\label{smelt}

In this section we consider the first order vortex lattice melting transition at
weak pinning strength.  Our methods for identifying the melting transition and establishing
that it is indeed a first order phase transition are the same as we have used in
our earlier work\cite{OTmelt} on the more dilute $f=1/20$ system.

\begin{figure}[tbp]
\epsfxsize=8.6truecm
\epsfbox{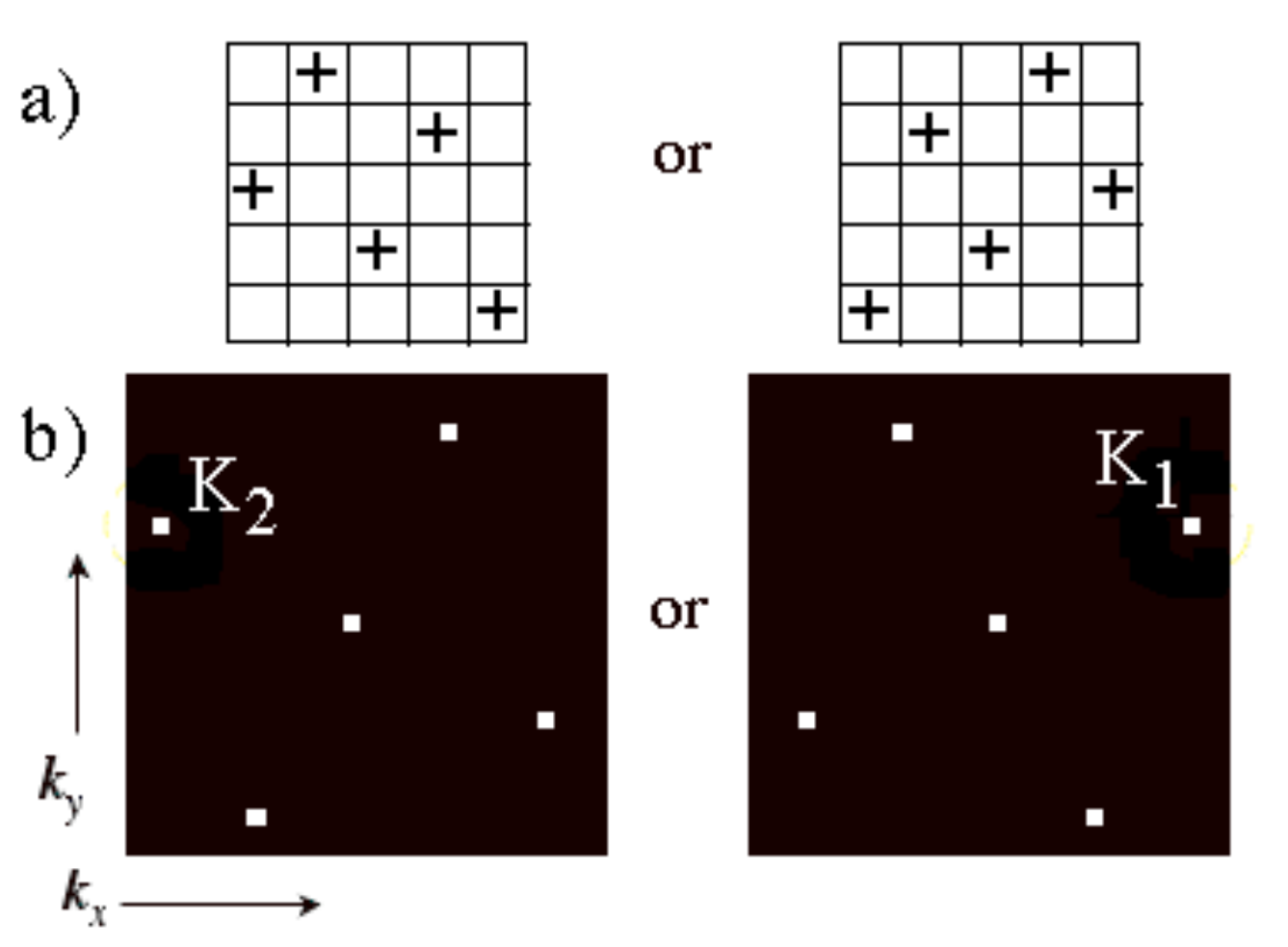}
\caption{
a) Real space configuration of vortex lines (denoted by ``$+$") piercing the $xy$ plane in the ground state of the pure ($p=0$) model, for a vortex line density $f=1/5$.  The two possible orientations of this ground state vortex lattice with respect to the underlying numerical grid are shown.
b) Intensity plot of vortex structure function $S({\bf k}_\perp)$ for each of the two ground state vortex lattice orientations shown in (a); ${\bf k}_\perp =0$ is at the center of the plot.  Peaks at the non-zero reciprocal lattice vectors are labeled ${\bf K}_1$ and ${\bf K}_2$ respectively for the two orientations.
}
\label{f2}
\end{figure}

For a pure system with disorder strength $p=0$ the vortex line lattice, of vortex
density $f=1/5$, will order into a ground state containing
a square vortex lattice with lattice constant $\sqrt{5}$.
There are two possible orientations of this square lattice with respect to the underlying grid,
related to each other by a reflection through the $\hat x$ axis, as shown in Fig.~\ref{f2}a.
In Fig.~\ref{f2}b we show the vortex structure function $S({\bf k}_\perp)$ for each
of these two ground state orientations.  We find sharp Bragg peaks at reciprocal
lattice vectors ${\bf K}$.  For the vortex lattice of Fig.~\ref{f2}a there are only 
four non-zero reciprocal lattice
vectors, related to one another by $\pi/2$ rotations of $k$-space.  For the two possible
ground state orientations, we label these two disjoint sets of reciprocal lattice vectors by ${\bf K}_1$
and ${\bf K}_2$ as shown in Fig.~\ref{f2}b.

At finite but small disorder strength, $p\le p_c\simeq 0.22$, we continue to find at
low temperatures a vortex lattice with the same symmetry as that of Fig.~\ref{f2}a.
The existence of the two possible orientations for this state motivates our definition
of the following order parameter for the vortex lattice to liquid melting transition.  If we denote
by $S({\bf K}_1)$ and $S({\bf K}_2)$ the value of the vortex structure function
averaged over  the four non-zero reciprocal lattice vectors of each orientation respectively,
then the difference, $\Delta S\equiv S({\bf K}_1)-S({\bf K}_2)$, will signal the vortex
lattice melting transition:  below melting, the system has ordered into one of the two
possible vortex lattice orientations, and so $S({\bf K}_1)$ is large and $S({\bf K}_2)$ is 
small, or vice versa, giving a large value of $|\Delta S|$; above melting, the system
is in a liquid state with the same symmetry as the underlying grid, so
$S({\bf K}_1)=S({\bf K}_2)$ by reflection symmetry and $\Delta S=0$.

\begin{figure}[tbp]
\epsfxsize=8.6truecm
\epsfbox{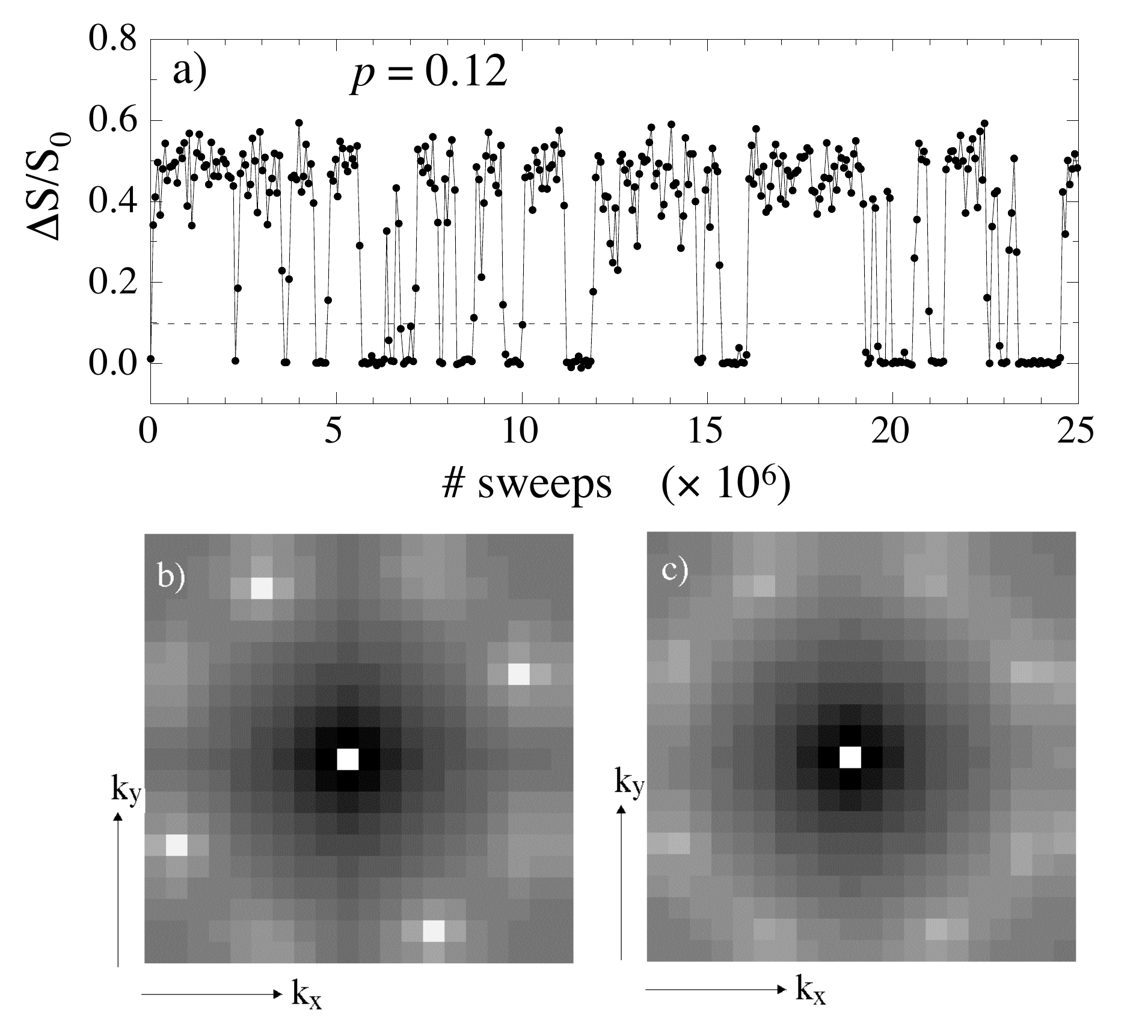}
\caption{
a) Plot of $\Delta S\equiv S({\bf K}_1)-S({\bf K}_2)$, normalized by $S_0\equiv S({\bf k}_\perp=0)$, vs Monte Carlo simulation time, close to the first order melting transition temperature $T_{\rm m}(p)\simeq 0.228$ for a particular realization of the quenched randomness  at disorder strength $p=0.12$.  The unit of time on the plot represents $10^6$ Monte Carlo sweeps through the entire system of grid size $L=20$, $L_z=6$, and each data point represents an average over $65536$ consecutive sweeps.  We see the system making many discrete transitions between vortex liquid, represented by small values of $\Delta S/S_0 \lesssim 0.1$, and vortex lattice, represented by large values of $\Delta S/S_0\gtrsim 0.1$. 
b) Logarithmic intensity plot of vortex structure function $\ln S({\bf k}_\perp)$ averaged over only the vortex {\it lattice} states of (a).  A sharp Bragg peak at ${\bf K}_1$ is seen.  c) Logarithmic intensity plot of $\ln S({\bf k}_\perp)$ averaged over only the vortex {\it liquid} states of (a).  Diffuse peaks of equal height are seen at both ${\bf K}_1$ and ${\bf K}_2$.
}
\label{f3}
\end{figure}

Anticipating a first order vortex lattice melting transition, we slowly cool down
a $20\times 20 \times 6$ size system from high temperature using ordinary Metropolis Monte Carlo until
we reach a temperature at which we observe the system to switch back and forth
between large and small values of $\Delta S$ during the course of the simulation.
As an example of this, we show in Fig.~\ref{f3}a a plot of $\Delta S/S_0$ vs. simulation time
($S_0\equiv S(0)=fL^2$) at a temperature close to the melting transition 
$T_{\rm m}(p)$ for a moderate value of disorder strength $p=0.12$.
In Fig.~\ref{f3}b we show an intensity plot of $\ln S({\bf k}_\perp)$, averaged over only those
configurations in Fig.~\ref{f3}a which have $\Delta S/S_0 >0.1$.  We see sharp Bragg peaks
of high intensity
at the reciprocal lattice vectors ${\bf K}_1$ indicating that this is the vortex lattice state.
In Fig.~\ref{f3}c we show show an intensity plot of $\ln S({\bf k}_\perp)$, averaged now over
only those configurations in Fig.~\ref{f3}a which have $\Delta S/S_0<0.1$.  We see broad
diffuse peaks of equal low intensity at both ${\bf K}_1$ and ${\bf K}_2$, indicating that this is
the vortex liquid state.

\begin{figure}[tbp]
\epsfxsize=8.6truecm
\epsfbox{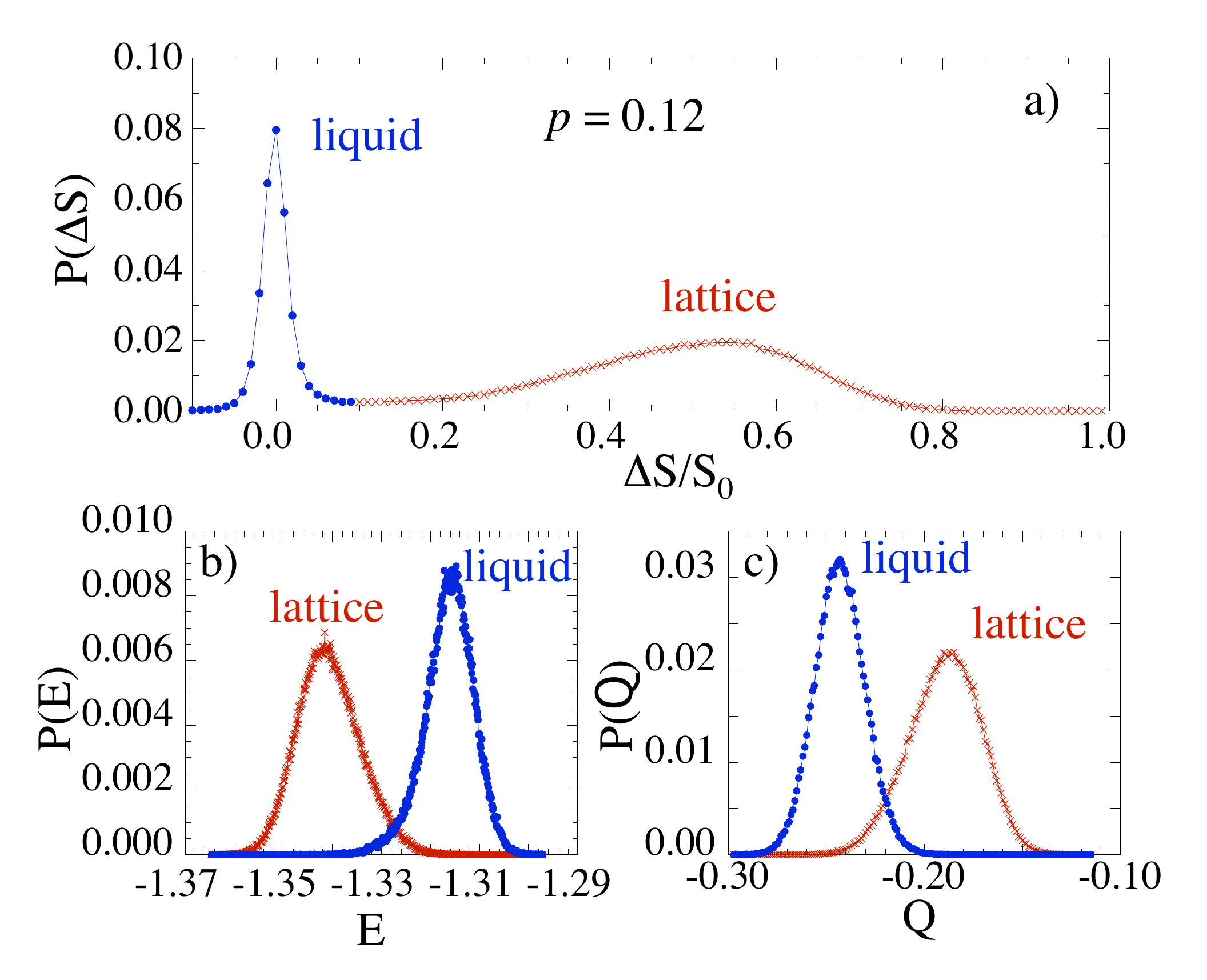}
\caption{(Color online)
a) Histogram $P(\Delta S)$ of values of $\Delta S/S_0$ obtained at the melting temperature
$T_{\rm m}(p)$ for a particular realization of the quenched randomness at disorder strength
$p=0.12$, system size $L=20$, $L_z=6$, from the data of Fig.~\ref{f3}a.  The minimum at $\Delta S/S_0\simeq	 0.1$ separates
configurations corresponding to the vortex liquid, represented by the peak centered at $\Delta S/S_0=0$, from the configurations corresponding to the vortex lattice, represented by the peak centered near $\Delta S/S_0\sim 0.55$.
b) Histograms $P(E)$ of energy per grid site $E$, and
c) histograms $P(Q)$ of the disorder conjugate variable $Q$,
for configurations in the vortex liquid phase ($\Delta S/S_0<0.1$) and the vortex lattice phase 
($\Delta S/S_0>0.1$) of (a).  Histograms indicate discontinuous jumps $\Delta E$ and $\Delta Q$ between vortex liquid and lattice at the melting transition.
}
\label{f4}
\end{figure}

In Fig.~\ref{f4}a we plot the histogram $P(\Delta S)$ of values of $\Delta S$ found in the data of Fig.~\ref{f3}a.
We see two well separated peaks centered at $\Delta S/S_0=0$ and $\Delta S/S_0\simeq 0.55$, representing the vortex
liquid and vortex lattice states respectively.  We define the melting transition temperature $T_{\rm m}(p)$
to be the temperature at which the area under these two peaks is equal.  In this manner, varying $p$ and averaging 
over $8$ independent realizations of the random disorder, we plot the vortex lattice melting line $T_{\rm m}(p)$ for
systems of size $20\times 20\times 6$ as the blue solid curve in Fig.~\ref{f1}.
As $p$ increases, $T_m(p)$ decreases while the slope $|dT_{\rm m}/dp|$ rapidly increases.
As $p$ increases towards $p_c\simeq 0.22$, with correspondingly low melting $T_{\rm m}$,
a failure to achieve proper equilibration of the system prevents us from continuing to 
trace out the melting curve to lower temperatures.

Next we demonstrate that, within our model system, melting remains a first order 
transition along the melting curve for as far as we can map it out.  There is no sign of 
it ending at an ``upper critical point" as has been often suggested by experimental works.\cite{Khaykovich,Nishizaki}
Choosing the minimum in the $P(\Delta S)$ histogram as the dividing point, we assign
each configuration as a vortex lattice or vortex liquid according to the value
of $\Delta S$ for that configuration.  Having divided configurations into 
distinct lattice and liquid states, we can then construct the histograms of 
energy, $P(E)$, and of the disorder conjugate variable, $P(Q)$, for each state respectively.
In Figs.~\ref{f4}b,c we show such histograms for disorder strength $p=0.12$,
corresponding to the data of Figs.~\ref{f3} and \ref{f4}a.  We see for both $E$ and
$Q$ well separated histograms for lattice and for liquid states.  Using these
histograms we then compute the average $E$ and $Q$ separately for the vortex lattice and
vortex liquid, and then compute the discontinuities in these quantities at melting,
\begin{eqnarray}
\Delta E &\equiv& \langle E\rangle_{\rm liquid} - \langle E\rangle_{\rm lattice}\\ \nonumber
\Delta Q &\equiv& \langle Q\rangle_{\rm lattice} - \langle Q\rangle_{\rm liquid}
\label{DEQ}
\end{eqnarray}

\begin{figure}[tbp]
\epsfxsize=8.6truecm
\epsfbox{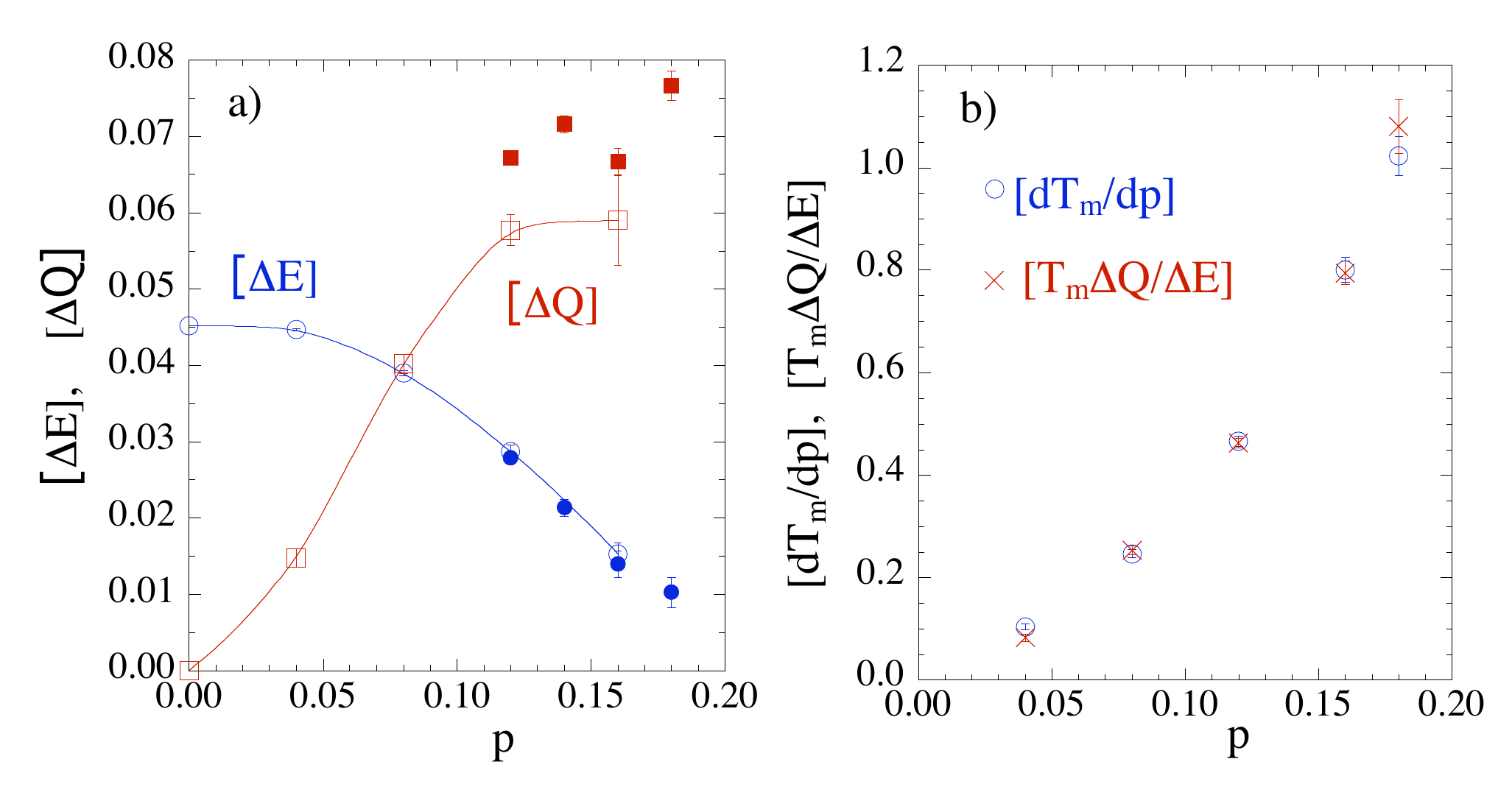}
\caption{(Color online)
a) open symbols: Disorder averaged discontinuities in the energy per site $[\Delta E]$ and disorder conjugate variable $[\Delta Q]$ at the melting transition $T_{\rm m}(p)$ vs disorder strength $p$, for systems of size $L=20$, $L_z=6$; solid symbols: average total changes $[\Delta E]$ and $[\Delta Q]$ in ordering from liquid to solid for systems of size $L=20$, $L_z=12$ (see discussion Sec.~\ref{sIW}).
While $[\Delta E]$ decreases with increasing $p$, $[\Delta Q]$ remains finite and grows, indicating that the melting transition remains strongly first order.
b) Comparison between the slope of the melting phase boundary $dT_{\rm m}/dp$ and the disorder averaged value of $[T_{\rm m}\Delta Q/\Delta E]$ vs disorder strength $p$, as a test
of the Clausius$-$Clapeyron relation, using data for system size $L=20$, $L_z=6$.
Good agreement indicates that our results are well equilibrated.
Error bars represent the estimated statistical error sampling over independent realizations
of the random disorder.
}
\label{f5}
\end{figure}

In Fig.~\ref{f5}a we plot (open symbols) the resulting $[\Delta E]$ and $[\Delta S]$, averaged over $8$ independent
realizations of the random disorder, vs disorder strength $p$, for the system of size $20\times20\times 6$.  As $p$ increases we see that $[\Delta E]$ decreases and appears to vanish.   A vanishing
$[\Delta E]$ implies a vanishing entropy jump and hence a vanishing of the delta-function
specific heat singularity usually associated with the first order melting transition.  
Such a vanishing of the specific heat singularity, observed experimentally on increasing 
applied magnetic field at fixed disorder strength, has been used as evidence for a weakening 
of the first order melting transition and its termination at a second order upper critical point.\cite{Bouquet}
However, if the first order phase transition is indeed to vanish at a second order critical point,
it is necessary that the discontinuities in {\it all} thermodynamic first derivatives of the free energy
vanish as the critical point is approached.  In Fig.~\ref{f5}a, however, we see that the 
discontinuity $[\Delta Q]$ {\it increases} as $p$ increases, and does not vanish as it must
if the first order line is to end in a critical point.  We thus find that our melting
transition remains strongly first order for all disorder strengths $p$.  The vanishing
of $[\Delta E]$ upon increasing $p$ merely reflects the increasing slope of the melting curve
$|dT_{\rm m}/dp|$ in accordance with the Clausisus-Clapeyron relation, as can be
seen as follows.  Since the free energies of lattice and liquid must be equal at $T_{\rm m}$,
\begin{equation}
\Delta{\cal F}[T_{\rm m}(p),p]\equiv {\cal F}_{\rm lattice}[T_{\rm m}(p),p]-{\cal F}_{\rm liquid}[T_{\rm m}(p),p]=0\enspace,
\end{equation}
we have,
\begin{equation}
{d\Delta{\cal F}\over dp}={\partial \Delta{\cal F}\over\partial p}+{\partial\Delta{\cal F}\over\partial T}{dT_{\rm m}\over dp} =0\enspace.
\label{eCC}
\end{equation}
Since
\begin{equation}
{\partial \Delta{\cal F}\over \partial p}= L^2L_z \Delta Q\quad {\rm and}\quad
{\partial \Delta{\cal F}\over \partial T}= L^2L_z {\Delta E\over T_{\rm m}}\enspace,
\end{equation}
substituting into Eq.~(\ref{eCC}) then gives the Clausius-Clapeyron relation for our system,
\begin{equation}
{dT_{\rm m}\over dp} = -{T_{\rm m}\Delta Q\over \Delta E}\enspace.
\label{eCC2}
\end{equation}
Fitting our data (blue solid circles in Fig.~\ref{f1}) for $T_{\rm m}(p)$ to a quadratic 
polynomial in $p^2$ (solid blue line in Fig.~\ref{f1}), we used the fitted polynomial to determine the slope $dT_{\rm m}/dp$,
and in Fig.~\ref{f5}b we plot $|dT_{\rm m}/dp|$ and the disorder averaged $[T_{\rm m}\Delta Q/\Delta E]$ vs disorder strength $p$.  The disorder average is over 8 independent realizations
of the quenched randomness.  We find excellent agreement with Eq.~(\ref{eCC2}), thus
verifying that our results are indeed very well equilibrated.

Experiment evidence for the {\it absence} of an upper critical point, in agreement with
our results, has been obtained in BSCCO by Avraham et al.\cite{Avraham} 
For an experimental system, in which disorder
strength is constant and the applied magnetic field $H$ is varied, the magnetization density
$M=(1/V)\partial{\cal F}/\partial H$ becomes the analog of our parameter $Q$.
While initial measurements of the jump $\Delta M$ at melting appeared to show
$\Delta M$ vanishing as $H$ increased, suggesting an upper critical point, subsequent
measurements using a additional small oscillating field to ``tickle" the vortex lines
to help avoid trapping in metastable local energy minima, showed a finite $\Delta M$
continuing along the melting curve past the presumed upper critical point and down to
even lower temperatures.  Their conclusion was that the presumed upper critical point
in BCSSO was an artifact of poor equilibration, and that a unified first order transition
line continued between thermally driven melting at low $H$, and disorder driven 
melting at larger $H$.  Similar conclusions had been drawn earlier by others\cite{vanderBeek, Gaifullin} based on measurements of the Josephson plasma frequency.

Finally, we consider the finite size dependence of our results to see that the values of
$[\Delta E]$ and $[\Delta Q]$ which we have found for the $20\times 20\times 6$ size
system do not appear to be decreasing (or perhaps vanishing) as the system size
increases.  The need to keep the transverse system length $L$ a multiple of 5, so
as to remain commensurate with the vortex lattice periodicity, and the difficulty
of equilibrating hops between lattice and liquid states as the system size, and hence
to the total free energy barrier between theses state, increases, limits greatly the range
of system sizes that we can consider.  In Table~\ref{tab1} we show our results
for the three system sizes $20\times 20\times 6$, $30\times 30\times 6$ and
$20\times 20\times 12$, for the specific disorder strength $p=0.12$.  We see that
while $T_{\rm m}$ decreases slightly as the system size increases, $[\Delta E]$
remains remain roughly independent of size while $[\Delta Q]$ shows a slight increase.
We also compute the spread in melting temperatures $\Delta T_{\rm m}$ that we
find as we consider different independent realizations of the quenched random
disorder $\{\epsilon_{i\mu}\}$.  If the system is self averaging over the quenched
disorder, we would expect that $\Delta T_{\rm m}\propto 1/\sqrt{V}$, with
$V=L^2L_z$ the system volume.  In Table~\ref{tab1} we therefore also give
the value for $\Delta T_{\rm m}\sqrt{V}$ for the three system sizes.  Considering the
relatively few (i.e. 8) disorder realizations we have considered for the two larger sizes,
and hence the corresponding large potential error in our estimate of $\Delta T_{\rm m}$,
we find our results consistent with the expectation of self averaging.

\begin{table}[tbp]
\caption{
Disorder averaged melting temperature $[T_{\rm m}]$ and discontinuities in the energy per site $[\Delta E]$ and disorder conjugate variable $[\Delta Q]$ for disorder strength $p=0.12$, for different system sizes $L^2\times L_z$.
The errors represent the estimated statistical error sampling over the number of independent
realizations of the random disorder specified in the last column.
$\Delta T_{\rm m}$ is the standard deviation of melting temperatures computed over the independent random realizations, and one expects $\Delta T_{\rm m}\propto 1/\sqrt{V}$, with $V=L^2L_z$ the volume of the system.  Results for the $20^2\times 12$ system are computed as described in Sec.~\ref{sIW}.
}
\begin{center}
\begin{tabular}{c|c|c|c|c|c}


$L^2\times L_z$ & $[T_{\rm m}]$ & $[\Delta E]$ & $[\Delta Q]$ &  $\Delta {T_{\rm m}}\sqrt{V}$ & \# \\ \hline \hline
$20^2\times 6$ & 0.2304 & 0.028 & 0.057 & 0.183 & 20 \\
         & $\pm$ 0.0009 & $\pm$ 0.0005 & $\pm$ 0.002 & & \\ \hline
$30^2\times 6$ & 0.227 & 0.026 & 0.061 & 0.190 & 8\\
         & $\pm$ 0.001 & $\pm$ 0.002 & $\pm$ 0.004 & & \\ \hline
$20^2\times 12$ & 0.225 & 0.0279 & 0.067 & 0.184 & 8 \\
         & $\pm$ 0.001 & $\pm$ 0.0004 & $\pm$  0.001 & & \\ 

\end{tabular}
\end{center}
\label{tab1}
\end{table}

\section{Vortex Glass at Strong Pinning}
\label{svg}

We now consider the strong pinning limit, $p>p_c\simeq 0.22$, and the possible existence of a vortex glass phase.  In our earlier work\cite{OTmelt} on the more dilute $f=1/20$ system, we presented preliminary 
evidence for the absence of a vortex glass phase within the XY model.  However later work by
Olsson\cite{OlssonVG} established that the model does indeed have a second order vortex glass transition at strong
disorder strength.  Moreover he found that critical scaling is isotropic.
Here we follow the analysis of Olsson, looking at the disorder averaged
helicity modulus in the $xy$ plane, $[\Upsilon_\perp]$, defined in Eq.~(\ref{eUps}).

\begin{figure}[tbp]
\epsfxsize=8.6truecm
\epsfbox{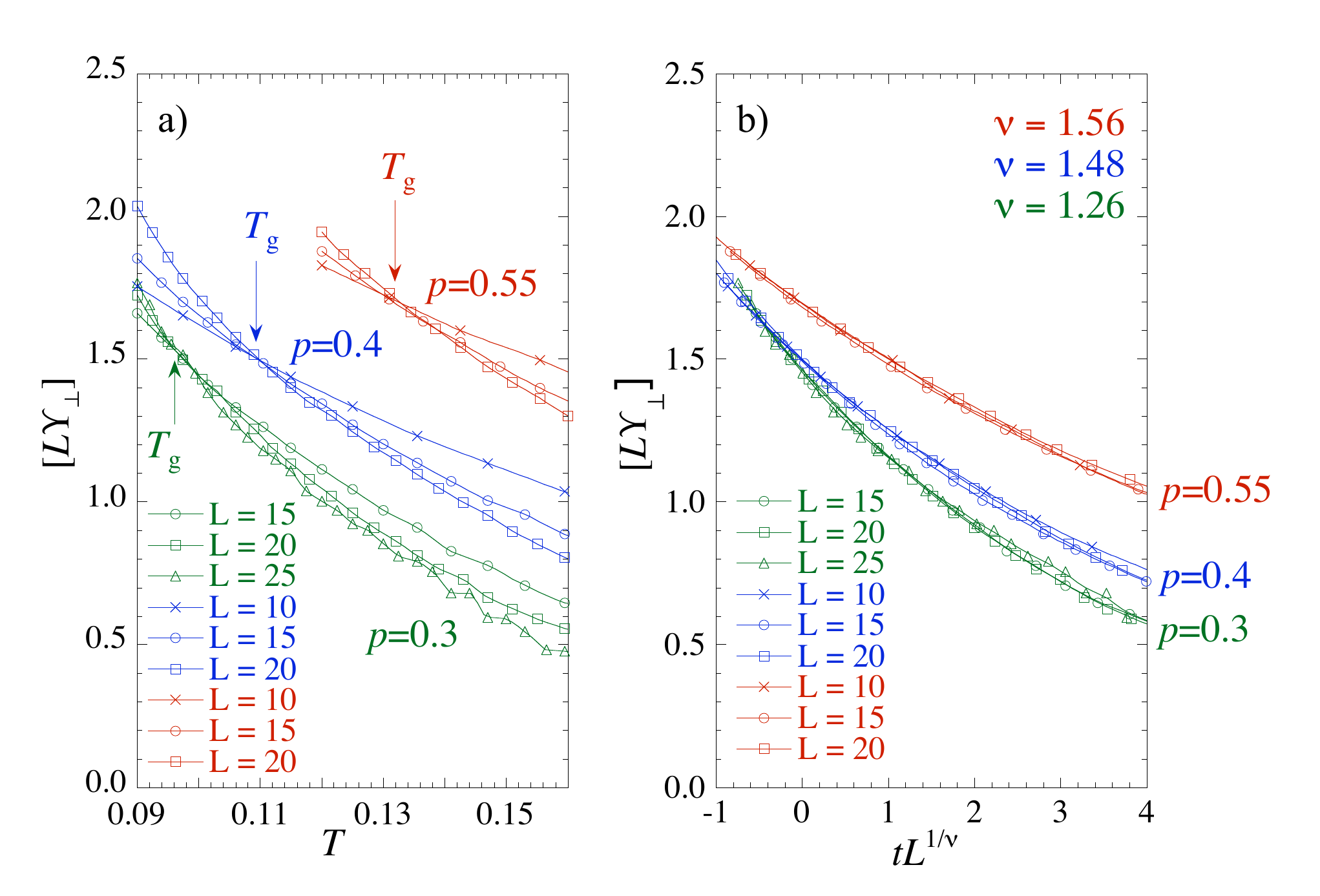}
\caption{(Color online)
a) Transverse helicity modulus scaled by system length $L\Upsilon_\perp$ vs temperature $T$
for different system sizes $L$, $L_z=(3/5)L$, for strong disorder strengths $p=0.30, 0.40$ and
$0.55$.  The common intersections of the curves for different $L$ locate the vortex glass transition temperatures $T_{\rm g}(p)$.  Results are averaged over $200$ independent realizations of the random disorder for the largest system size, and between $300-600$ realizations for smaller sizes.
b) $L\Upsilon_\perp$ of (a) replotted vs scaled temperature $tL^{1/\nu}$, with $t\equiv T-T_g$.
Collapse of the data for different $L$ to a common scaling curve determines the correlation length critical exponent $\nu\sim 1.5$.
}
\label{f6}
\end{figure}

In Fig.~\ref{f6}a we plot $[L\Upsilon_\perp]$ vs $T$ for three different disorder strengths,
$p=0.3, 0.4$ and $0.55$, using for each case three different system sizes from $L=10$ to $25$,
$L_z=(3/5)L$, as indicated in the figure.  Results are averaged over $200-600$ independent realizations
of the random disorder, depending on system size.  As discussed following Eq.~(\ref{eUpsScale}), for each value of $p$, the common intersection of the curves for different $L$ locates the vortex glass transition, $T_{\rm g}(p)$, and thus allows us to map out the vortex glass transition line, 
plotted as the black squares and black solid line in Fig.~\ref{f1}.  We see that $T_{\rm g}(p)$ increases for increasing $p$.    

In Fig.~\ref{f6}b we replot our data vs the scaled temperature $tL^{1/\nu}$, where the critical 
exponent $\nu$ has been chosen for each $p$ so as to give the best data collapse to
a common scaling curve for the different system sizes $L$, in accordance with the scaling 
equation of Eq.~(\ref{eUpsScale}).  For the two largest values $p=0.4$ and $0.55$ we
find $\nu\sim 1.5$, in agreement with the more precise calculations of Olsson.
For the smallest $p=0.3$, we find a smaller $\nu\sim 1.3$.  We believe that for this last
case, where our data (see Fig.~\ref{f6}a) is the
noisiest  and as $T_{\rm g}$ is the smallest we have the least data at $T<T_{\rm g}$, the fact 
that we are closest to the vortex lattice melting transition may mean that our system
sizes are still too small and so we are in a cross-over region rather than the true large $L$ scaling limit.

The existence of a vortex glass phase in the bond disordered 3D uniformly frustrated
XY model has also been claimed in simulations by Kawamura,\cite{Kawamura0, Kawamura} who in addition to 
measuring a value $\nu=1.1\pm 0.2$ also measures the critical exponent $\eta=-0.5\pm 0.1$.
These are also consistent with the values $\nu= 1.3\pm 0.2$ and $\eta= -0.4\pm  0.1$ found by Lidmar\cite{Lidmar} using an elastic model of interacting vortex lines that includes dislocations.
These values for the critical exponents lie in reasonable agreement with the
values $\nu_{\rm GG}=1.39\pm0.20$ and $\eta_{\rm GG}=-0.47\pm 0.07$ found for the 
simpler 3D {\it gauge glass} model.\cite{Katzgraber}  The gauge glass has the same Hamiltonian as Eq.~(\ref{eH}),
except that the bond couplings $J_{i\mu}$ are taken as uniform, and the randomness is put
in the vector potential, with $A_{i\mu}$ chosen randomly from a uniform distribution on
$(-\pi,\pi]$.  In the gauge glass, the average magnetic field thus vanishes in all directions and
so the model is intrinsically isotropic.  This comparison of critical exponents suggests that the vortex glass and the gauge glass may be in the same universality class.  If so, Kawamura notes\cite{Kawamura} that
the addition of magnetic field fluctuations associated with a finite magnetic
penetration length $\lambda$ (``magnetic screening") in a real superconductor
may serve to destabilize the vortex glass transition and replace it with a smooth crossover behavior,
as has been numerically observed to happen in the gauge glass\cite{Bokil} (such an effect was observed
in early simulation by Kawamura\cite{Kawamura0} of the vortex glass with magnetic screening, however the
use of free boundary conditions and relatively small system sizes in that work 
raise questions concerning its validity).  If this scenario is correct, it suggests that the vortex glass transition line we find becomes only a crossover phenomenon in a real physical superconductor with finite $\lambda$.  
This crossover might still be observed in current-voltage characteristics that seem to
obey critical scaling, as at a second order glass transition, only to have the scaling break down
at sufficiently small currents (which probe increasingly large length scales and thus are 
sensitive to finite $\lambda$).  The failure to conclusively demonstrate critical scaling in experimental data that has been noted by Strachan et al,\cite{Strachan} may perhaps be due to such an effect.
In contrast to this situation with respect to the vortex glass, we note that magnetic field fluctuations should not qualitatively 
effect the vortex lattice melting transition at low disorder strength, since melting is believed to be
mediated by short wavelength shear fluctuations and so should be relatively insensitive to 
behavior on the long length scale $\lambda$.

Looking at the vortex glass $T_{\rm g}(p)$ and vortex lattice melting $T_{\rm m}(p)$
transition lines in the $T-p$ plane, shown in Fig.~\ref{f1}, it appears that these lines are
of distinctly different origin, rather than one being a continuation of the other.  However
we are unable to directly investigate this issue as we are unable to sufficiently equilibrate 
our model system at the low temperatures where these two transition lines would appear to meet.

\section{Smeared Melting at Intermediate Pinning}
\label{sIW}

\begin{figure}[tbp]
\epsfxsize=8.6truecm
\epsfbox{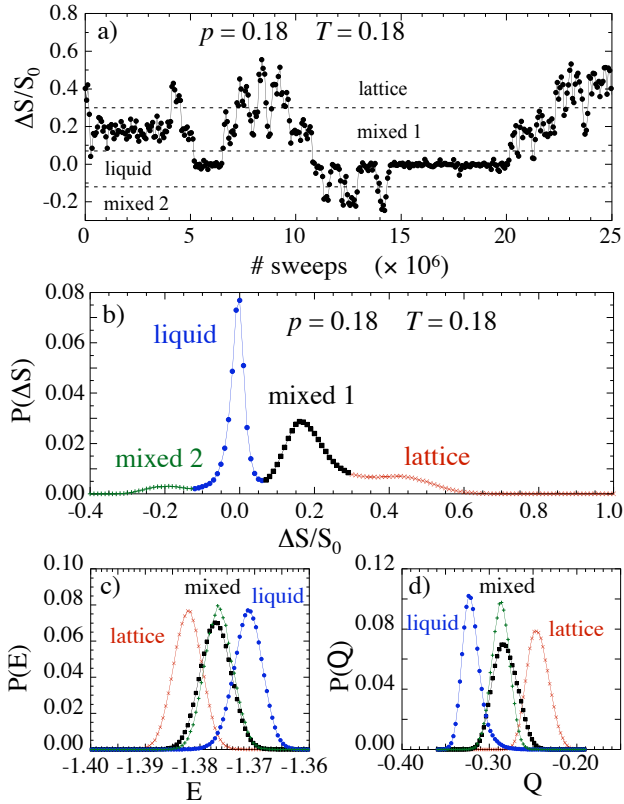}
\caption{(Color online)
a) Plot of $\Delta S\equiv S({\bf K}_1)-S({\bf K}_2)$, normalized by $S_0\equiv S({\bf k}_\perp=0)$, vs Monte Carlo simulation time, at $T=0.18$ and disorder strength $p=0.18$, for a particular realization of the quenched randomness  in a system of size $20\times 20\times 12$.  The unit of time on the plot represents $10^6$ Monte Carlo sweeps through the system, and each data point represents an average over $65536$ consecutive sweeps.  The system shows discrete transitions between states with four different average values of $\Delta S$; these states are separated by the horizontal dashed lines and labeled ``lattice", ``liquid", ``mixed 1" and ``mixed 2". b) Histogram $P(\Delta S)$ vs $\Delta S/S_0$ for the data of part a, showing four peaks corresponding to the four different states. c,d) Histograms of energy $E$ and disorder conjugate variable $Q$ for each of the four states.
}
\label{f7}
\end{figure}

We return now to the vortex lattice melting at $p<p_c\simeq 0.22$.  Considering
a system size $20\times 20\times 12$, twice as thick as the size considered in section \ref{smelt},
we find that as $p$ increases above $\sim 0.12$ the sharp first order melting transition
that we observed in the thinner system now smears out over a finite temperature interval,
$T_{\rm m2}\lesssim T\lesssim T_{\rm m1}$.  In Fig.~\ref{f7}a we show a plot of the ordering parameter $\Delta S/S_0$ vs simulation time for a particular realization of the quenched random disorder at disorder strength $p=0.18$ and temperature $T_{\rm m2}<T=0.18 <T_{\rm m1}$.  In contrast to what we found in Fig.~\ref{f3} for the thinner system at weaker disorder, we now see that $\Delta S/S_0$ appears to fluctuate about {\it four} different discrete values.  In Fig.~\ref{f7}b we plot the histogram $P(\Delta S)$ vs $\Delta S/S_0$ resulting from the data of Fig.~\ref{f7}a.  We see clearly four distinct peaks which we identify as representing the lattice state ($\Delta S/S_0 \sim 0.45$), the liquid state ($\Delta S/S_0\sim 0$), and what we will denote as the ``mixed 1" and ``mixed 2" states ($\Delta S/S_0\sim\pm 0.2$).  The locations of the minima between these peaks, shown as the dashed lines in Fig.~\ref{f7}a, we use as a simple criteria for sorting each of the microscopic configurations into one of these four different states.  Having so sorted the microscopic states we can then compute the histograms of energy $E$ and disorder conjugate variable $Q$ for each of the four states.  These we show in Figs.~\ref{f7}c and d.  We see that there is a discontinuous decrease in $E$, and a discontinuous increase in  $Q$, as the system transitions from the liquid, to the mixed state, to the lattice.  The distributions of $E$ and $Q$ are comparable for both mixed 1 and mixed 2 states.

\begin{figure}[tbp]
\epsfxsize=8.6truecm
\epsfbox{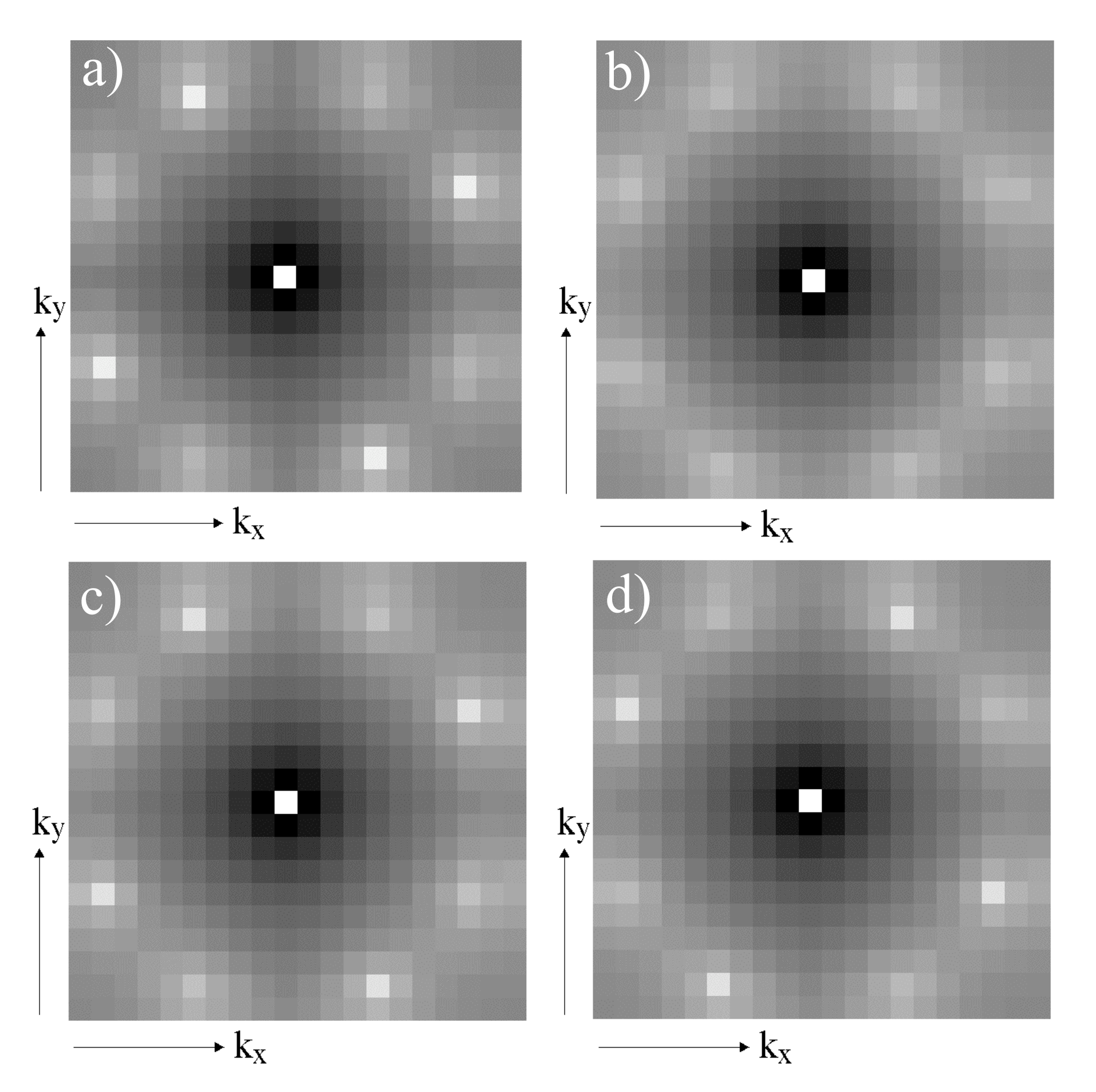}
\caption{
Logarithmic intensity plot of vortex structure function $\ln S({\bf k}_\perp)$ averaged over the microscopic configurations of the (a) lattice, (b) liquid, (c) mixed 1 and (d) mixed 2 states, corresponding to the data of Fig.~\protect\ref{f7}.}
\label{f8}
\end{figure}

In Fig.~\ref{f8} we show log intensity plots of the structure function, $\ln S({\bf k}_\perp)$, averaged separately over the configurations belonging to each of the four states.  Figs.~\ref{f8}a$-$d correspond to the lattice, liquid, mixed 1 and mixed 2 states, respectively.  We see that the lattice state, as expected, has sharp Bragg peaks of relatively high intensity about the reciprocal lattice vectors ${\bf K}_1$.  The liquid state has only broad diffuse peaks of equal low intensity at both ${\bf K}_1$ and ${\bf K}_2$  (see Fig.~\ref{f2}b for the definition of ${\bf K}_1$ and ${\bf K}_2$).  The mixed 1 and mixed 2 states have sharp peaks of intermediate intensity at  ${\bf K}_1$ and ${\bf K}_2$ respectively.

\begin{figure}[tbp]
\epsfxsize=8.6truecm
\epsfbox{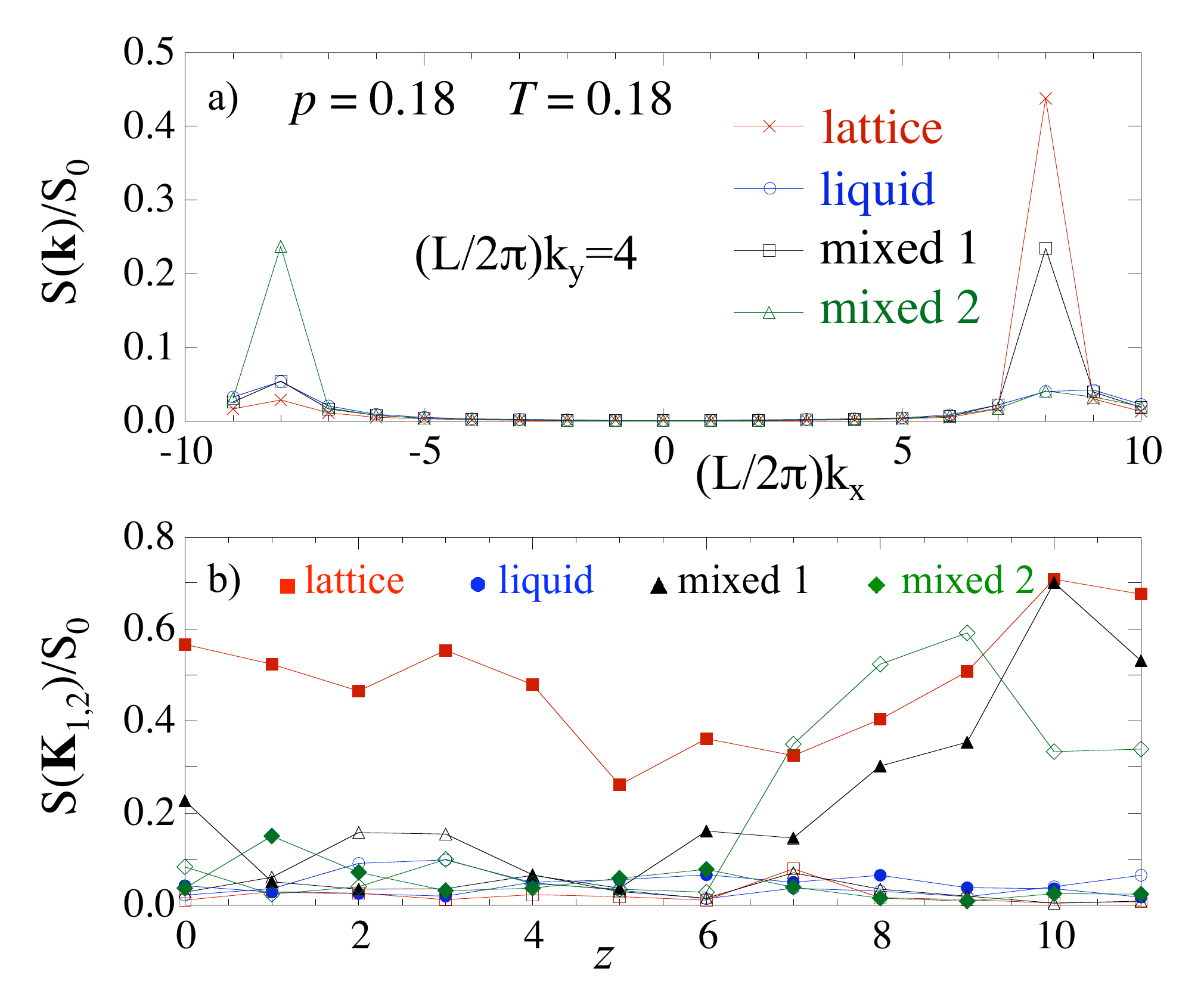}
\caption{(Color online)
a) Plot of vortex structure function $S(k_x,k_y)/S_0$ vs $k_x$, for fixed $k_y=4(2\pi/L)$ passing through the reciprocal lattice vectors ${\bf K}_1$ and ${\bf K}_2$, for the lattice, liquid, mixed 1 and mixed 2 states, whose intensity plots are shown in Figs.~\ref{f8}a$-$d.
b)Plot of $S({\bf K}_1)$ (solid symbols) and $S({\bf K}_2)$ (open symbols), computed for the individual layer at height $z$, vs $z$.  Results are shown for the lattice, liquid, mixed 1 and mixed 2 states of Fig.~\ref{f8}.
}
\label{f9}
\end{figure}

To investigate the nature of the mixed states, we plot in Fig.~\ref{f9}a the profile of $S(k_x,k_y)$ vs $k_x$ for fixed $k_y=4(2\pi/L)$ passing through the wavevectors  ${\bf K}_1$ and ${\bf K}_2$.  We see that the peak at ${\bf K}_1$ for the mixed 1 state of Fig.~\ref{f8}c is as sharp as, but smaller in amplitude than, the Bragg peak of the vortex lattice state of Fig.~\ref{f8}a, while the remainder of this profile (in particular the small diffuse peak at ${\bf K}_2$) resembles that of the liquid state of Fig.~\ref{f8}b.  A similar result holds for the mixed 2 state of Fig.~\ref{f8}d, expect that the behaviors at ${\bf K}_1$ and ${\bf K}_2$ are now interchanged.  In Fig.~\ref{f9}b we plot the values of $S({\bf K}_1)$ and $S({\bf K}_2)$, computed now for individual $xy$ planes (instead of averaged over all such planes as in Eq.~(\ref{eSk})), vs the plane height $z$.  We see that in the mixed 1,2 states roughly half the system is ordered into a lattice, while half appears disordered.  Comparing the plots for mixed 1 and mixed 2, we see that the location of the more strongly ordered planes remains roughly the same, independent of the two possible orientations of the vortex lattice that is ordering in those planes.  We therefore conclude that the relative order or disorder of planes arises from the specific distribution of random bond strengths $\{J_{i\mu}\}$ that exists in the particular realization of the quenched randomness, rather than being an effect due to thermal fluctuations or being stuck out of equilibrium.

\begin{figure}[tbp]
\epsfxsize=8.6truecm
\epsfbox{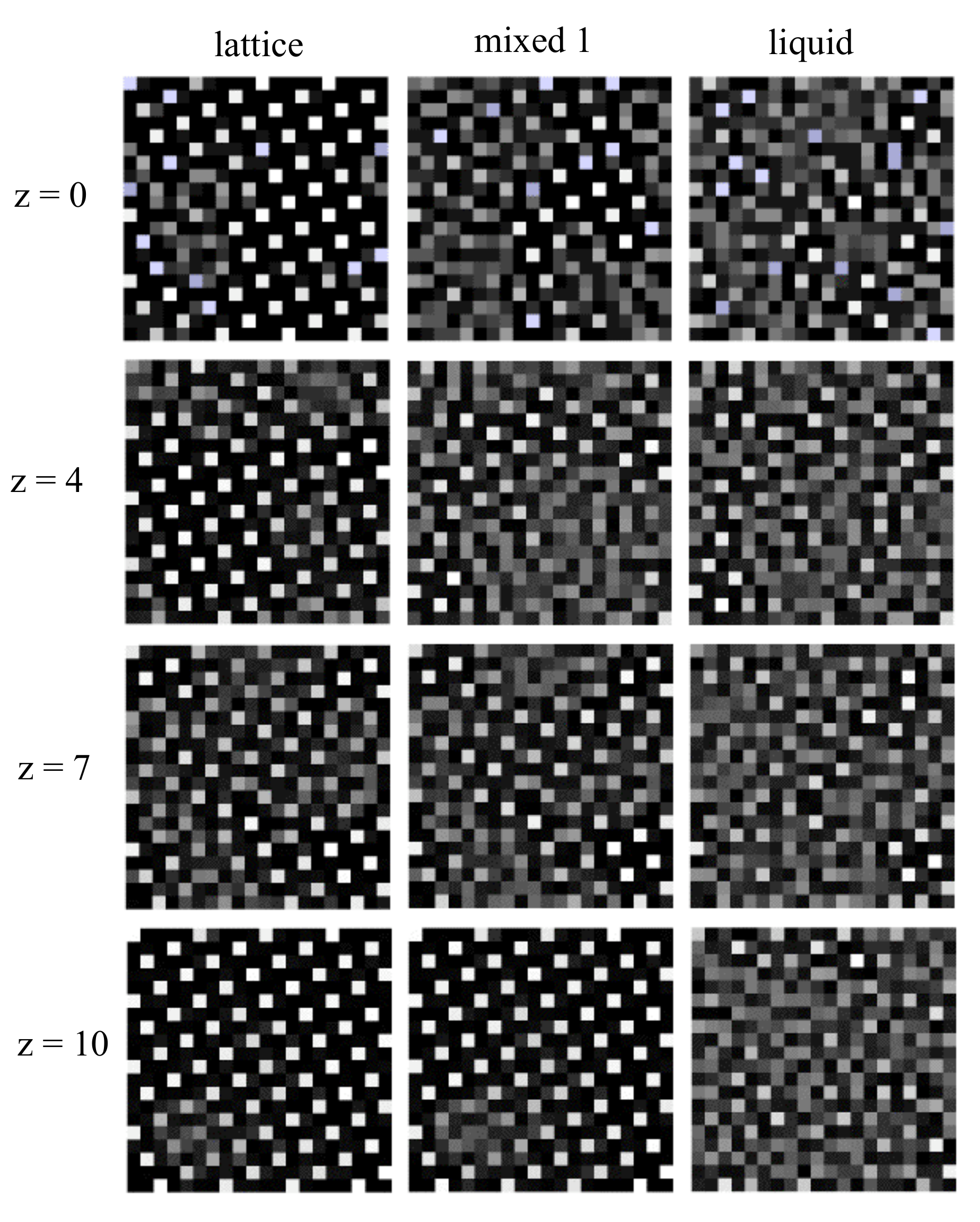}
\caption{
Intensity plots of real space average vorticity $\langle n_z(x,y,z)\rangle$ in the $xy$ plane at four representative layers, $z=0$, $4$, $7$, and $10$, for the lattice, mixed 1, and liquid states, corresponding to the data of Fig.~\ref{f8}.
}
\label{f10}
\end{figure}

To further understand the nature of the mixed state, in Fig.~\ref{f10} we show intensity plots of the real space average vorticity $\langle n_z(x,y,z)\rangle$ in the $xy$ plane for several representative layers at heights $z$.  We show plots for the lattice, mixed 1, and liquid states, for the same run that produced the results shown in Figs.~\ref{f7}$-$\ref{f9}.  A bright white square represents a pinned vortex that stays on that site throughout the course of the simulation run; a dark black square represents a site on which no vortex sits; squares of intermediate shades of gray represent sites that vortices hop into and out of during the course of the simulation.  Hence a region of periodic white squares indicates a local region that has ordered into a pinned vortex lattice; a region of gray squares with no apparent structure to the relative shadings indicates a local region that is disordered into a vortex liquid.  We see from Fig.~\ref{f10} that even the state which is globally characterized as the lattice contains localized regions which are disordered;  these are presumably regions where the vortex pinning, due to the
local bond disorder $\{J_{i\mu}\}$, is stronger than average.  In one layer, $z=7$, we see large domains of each of the two different possible vortex lattice orientations (i.e. a domain with ordering wavevector ${\bf K}_2$, as well as a domain with the dominant ordering wavevector ${\bf K}_1$).  In the state which is globally characterized as a vortex liquid, we still see individually pinned vortices.  In the mixed state we see layers that are mostly lattice ($z=10$), layers that are mostly liquid ($z=4$), and layers that consist of coexisting domains of lattice and liquid ($z=0,7$).

\begin{figure}[tbp]
\epsfxsize=8.6truecm
\epsfbox{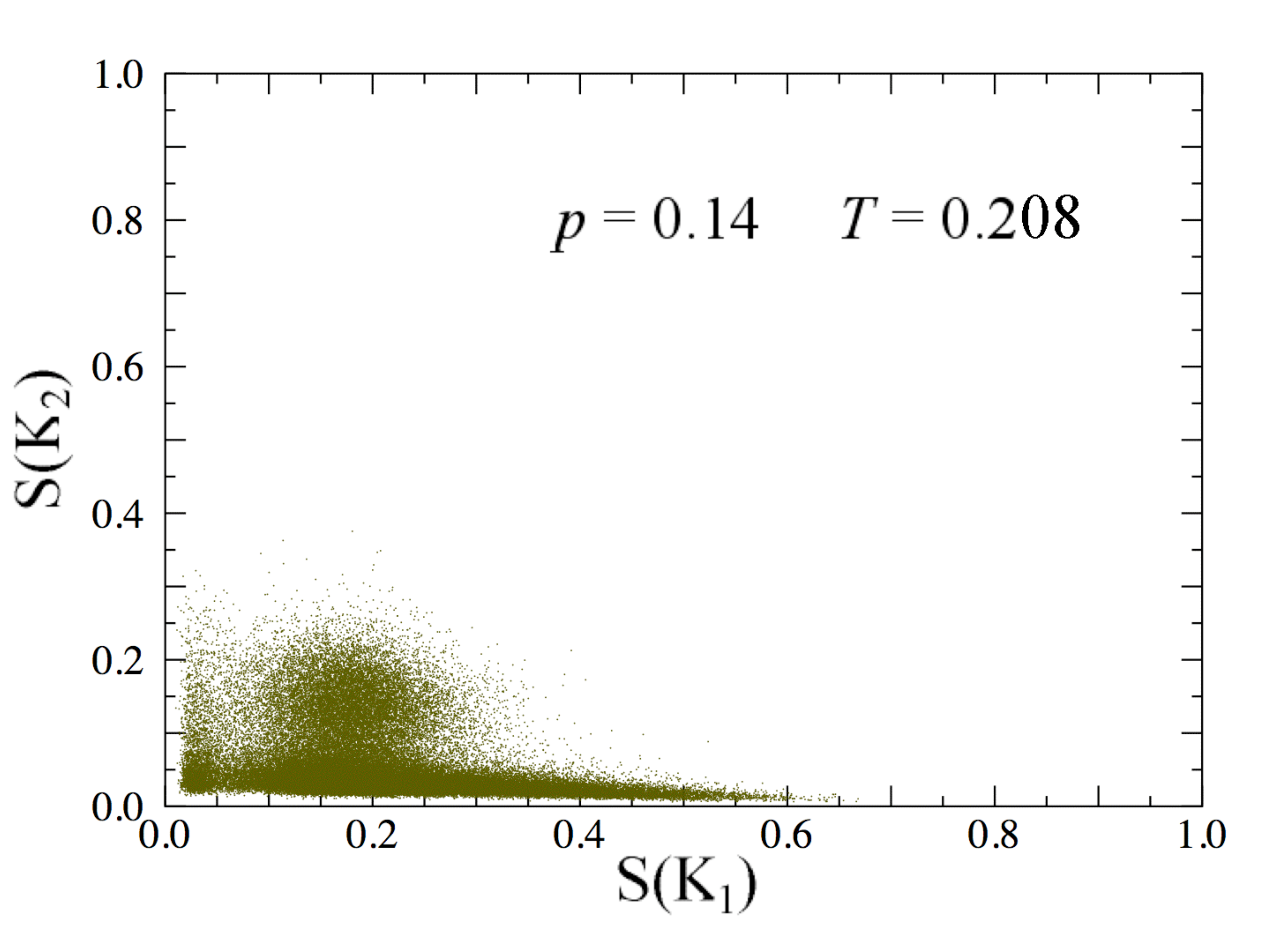}
\caption{
Intensity plot of two dimensional histogram $P(S({\bf K}_1, {\bf K}_2))$ for a particular realization of the random disorder at $p=0.14$, $T=0.208$.
}
\label{f11}
\end{figure}

The above results suggest that the mixed state is one in which different regions of the system order from liquid to lattice at slightly different temperatures, due to fluctuations in local bond strength disorder.  Depending on the particular realization of disorder, this can happen in ways that are more complex than the particular case illustrated in Figs.~\ref{f7}$-$\ref{f10}.  In some other realizations we have found the histogram $P(\Delta S)$ does not show such clearly separated peaks as in Fig.~\ref{f7}b, and there may be suggestions of more than three peaks.  In some realizations we find a mixed state in which some layers of the system have a lattice ordering specified by the wavevector ${\bf K}_1$, while other layers of the system are ordered with the opposite lattice orientation specified by the wavevector ${\bf K}_2$.  For such a case our ordering parameter $\Delta S=S({\bf K}_1)-S({\bf K}_2)$ can be very small, which by our previous criterion would misidentify such a state as a liquid.  To avoid such misidentifications, we adopt instead a more general approach, looking at the two dimensional histogram $P(S({\bf K}_1), S({\bf K}_2))$.  In such a histogram, the liquid state is represented by a peak at $S({\bf K}_1)=S({\bf K}_2)=0$, a uniformly ordered lattice state is represented by a peak at either $S({\bf K}_1)$ large {\it and} $S({\bf K}_2)$ small, or vice versa, and a mixed state is represented by a peak elsewhere in the $S({\bf K}_1)\times S({\bf K}_2)$ plane.  In Fig.~\ref{f11} we show such a two dimensional histogram for a different particular realization of the randomness at disorder strength $p=0.14$ and temperature $T=0.208$.  We see the lattice state with wavevector ${\bf K}_1$ coexisting with the liquid state, coexisting with a mixed state having roughly equal ordering in ${\bf K}_1$ and ${\bf K}_2$.
Based on the location of the peaks in this two dimensional $P(S({\bf K}_1), S({\bf K}_2))$ histogram we adopt the following criteria for sorting microscopic configurations into liquid, lattice and mixed states: if $S({\bf K}_1)/S_0 < 0.08$ {\it and} $S({\bf K}_2)/S_0<0.08$ we assume the configuration is a liquid; if $S({\bf K}_1)/S_0 > 0.3$ {\it and} $S({\bf K}_2)/S_0 < 0.1$ or vice versa, we assume the configuration is a uniform lattice; any other configuration is taken as belonging to the mixed state.

As temperature is varied, the locations of the peaks in  $P(S({\bf K}_1), S({\bf K}_2))$ vary only slightly, however their weights (total number of configurations in each peak)  are found to vary in the following manner.  As $T$ decreases, the weight of the liquid state decreases while the weights of the other states increase.  As $T$ is decreased further, the liquid state disappears, the weight for the mixed state decreases, while the weight for the lattice state increases.  Finally, at low enough $T$, the mixed state disappears and only the lattice state remains.  We define $T_{\rm m1}$ as the temperature where the liquid state and mixed state are roughly equal in weight.  We define $T_{\rm m2}$ as the temperature where the lattice state and mixed state are roughly equal in weight.  Our determinations of $T_{\rm m1}$ and $T_{\rm m2}$ in this manner are made from rough eyeball estimates, as we find that the errors involved in such eyeball estimates are considerably smaller than the the sample to sample variation in these temperatures between different independent realizations of the quenched random disorder.  These values of $T_{\rm m1}$ and $T_{\rm m2}$, averaged over $8$ independent realizations of the quenched randomness, determine the phase boundaries shown in Fig.~\ref{f1}.  
Choosing a temperature $T_{\rm m2}<T<T_{\rm m1}$ where both liquid and lattice coexist, we also compute
the averages of $E$ and $Q$ in the liquid and lattice states separately.   We then compute $\Delta E$ and $\Delta Q$ as in Eq.~(\ref{DEQ}), representing the total change in these quantities as the system orders from liquid to lattice.  For a few random realizations, where our simulation temperatures did not include a value where liquid and lattice coexisted, we used standard extrapolation methods to extrapolate the liquid (lattice) histogram from a slightly higher (lower) temperature to a common temperature $T_{\rm m2}<T<T_{\rm m1}$, and then computed $\Delta E$ and $\Delta Q$ from these extrapolated histograms.  Our results,
averaged over $8$ independent realization of the quenched random disorder,  are shown in Fig.~\ref{f5}a (solid symbols) where we see that they appear to be a consistent continuation of the results found in the smaller size system ($20\times 20\times 6$) at weaker disorder strengths $p$, except that, as noted previously in Table~\ref{tab1}, $[\Delta Q]$ is somewhat larger in the larger system size.

The above results suggest that as the system size and the disorder strength increase, the transition from vortex liquid to lattice remains {\it locally} first-order-like, as was found in Section \ref{smelt}, however different regions of the system may order at slightly different temperatures.  Such a conclusion is in perfect agreement with experiment results from magneto-optical imaging of vortex lattice melting in BSCCO,\cite{Soibel} where it was observed that the spread in local melting temperatures over the area of the sample increased as the average melting temperature decreased (i.e. as the average pinning strength increased).  A similar observation of coexisting ordered and disordered vortex regions over a finite temperature interval at the vortex lattice melting transition was reported in more recent experimental studies\cite{Pasquini} of the peak effect region of NbSe$_2$.

The vortex lattice melting transition thus becomes smeared out over a range of temperatures when viewed on the {\it global} scale.  Such a scenario was predicted long ago by Imry and Wortis,\cite{IW} who considered the effect of quenched randomness on a first order phase transition in terms of a competition between the decrease of the free energy due to local ordering of domains of finite size in regions of lower than average quenched disorder, vs the increase of free energy due the surface tension of the resulting domain walls enclosing the ordered domains.  We have attempted a quantitative test of the Imry-Wortis scenario as applied to our model.  Our efforts are described in Appendix A.  While they are suggestive, showing the right trends as disorder strength $p$ increases, our results are not conclusive.  Even if the Imry-Wortis scenario applies to our mixed state, two possibilities still exist: ({\it i}) it may be that on the global scale in the thermodynamic limit, there remains a sharp first order phase transition with reduced but still discontinuous jumps $\Delta E$ and $\Delta Q$ at a single well defined $T_{\rm m}$, or ({\it ii}) it may be that there  will be multitude of local transitions spread out smoothly over a range of spatially varying local melting temperatures, thereby converting the transition on the global long length scale to a continuous second order transition.  To discriminate between these two possibilities would require looking at much larger systems sizes than we are currently able to equilibrate.

\section{Discussion}

To summarize, our results for the vortex phase diagram of the uniformly frustrated 3D XY model with disordered couplings show many qualitative features in good agreement with experiments on strong type-II superconductors.  We find a sharp local first order vortex lattice melting transition at low pinning strengths $p$.  The melting temperature decreases as $p$ increases, and appears to be steadily decreasing towards zero as a critical pinning strength $p_c$ is approached.  We find that this melting transition remains first order, with no evidence for a critical end point or other multicritical points, down to the lowest temperatures to which we can equilibrate.  At high pinning strengths $p>p_c$ we find in our model a sharp second order vortex glass transition.  This glass transition may evolve into a non-signular crossover phenomenon if magnetic field fluctuations, due to a finite magnetic penetration length $\lambda$, were incorporated into the model.  A completely new feature of our simulations is the observation that at intermediate disorder strengths $p\lesssim p_c$ the vortex lattice melting transition is smeared out over a temperature interval of finite width, corresponding to coexisting regions of ordered and disordered vortex states, as has been seen in recent experiments.\cite{Soibel,Pasquini}

It is interesting to compare our results to those of Nonomura and Hu\cite{NonoHu} (NH), who studied a very similar 3D XY model but with much more weakly coupled planes, $J_z/J_\perp=1/400$, a more dilute vortex density $f=1/25$, and with random couplings that model a dilute set of localized strong pins rather than the amorphous random couplings we have used here. 
In addition to vortex lattice melting and vortex glass transitions, NH reported the existence of a vortex slush phase, lying between the vortex liquid and the vortex glass, and  separated from the vortex liquid by a sharp first order transition.  In earlier work\cite{OTcomment} we have repeated simulations of NH's model, using the exact same parameters as NH.   We find that their vortex slush phase shares some similarities with our intermediate region discussed above, in that both are regions in which the vortex lines have only partially ordered.  There are however some important differences.  

(i) Our intermediate region lies between the vortex liquid and vortex lattice phases and so may be thought of as a broadening of the melting transition; when we cool at fixed $p$ though the intermediate region, our system (except in rare cases when we fail to equilibrate) always orders into a clear vortex lattice.  In NH's model, the vortex slush lies above the vortex lattice phase (similar to what was reported in some experiments\cite{Worthington, Shibata}); cooling through NH's vortex slush, one enters the vortex glass and not the vortex lattice.  (ii) We showed\cite{OTcomment} that considerable hysteresis existed  in the region of NH's vortex slush phase.  In our present simulations there is no hysteresis: at fixed simulation parameters in our intermediate region the system is repeatedly hopping into and out of the vortex lattice, liquid, and mixed states (see Fig.~\ref{f7}a) and our system is thus fully equilibrated. (iii) We argued\cite{OTcomment} that in NH's model, most planes of their vortex slush contained an ordered vortex lattice, however the orientations of the vortex lattice varied with height $z$; we argued that these mismatched vortex lattice orientations would be unfavorable in the thermodynamic limit and that NH's vortex slush was most likely a finite size effect to be replaced by an ordered vortex lattice as system size increased.  In our intermediate region, however, we see coexisting planes of mostly vortex lattice, planes of mostly vortex liquid, as well as planes with large domains of both lattice and liquid.  The scaling argument we used against  NH's vortex slush thus does not apply.  It is of course possible that upon increasing system size, NH's vortex slush will similarly develop coexisting ordered and disordered domains within individual planes and so remain as a stable phase.  

We conclude therefore that our intermediate region is distinctly different from the vortex slush of NH, and is perhaps more similar to the multidomain glass state that has been proposed by Menon.\cite{Menon}  Unfortuntely, we have not been able to equilibrate our system in the interesting region where the vortex lattice melting transition meets the vortex glass transition.

Another set of interesting simulations has been carried out recently by Dasgupta and Valls\cite{DasguptaValls1,DasguptaValls2}  using a density functional approach applied to interacting pancake vortices in a 3D layered system.  They consider both the case of dense amorphous pins,\cite{DasguptaValls1} such as we consider here, and the case of dilute well localized pins,\cite{DasguptaValls2} closer to the model of NH, mapping out the phase diagram as a function of temperature and pinning strength.  Their approach, being essentially an equilibrium mean field method in the presence of quenched randomness,  is suited to locating first order phase transitions, as in vortex lattice melting or the proposed vortex slush,  rather than continuous second order transitions, as one expects for a vortex glass. For the dense amorphous pinning, they find only a single unified vortex lattice melting transition, qualitatively similar in shape to what we find in the present work.  Their vortex liquid state shows no significant local ordering on length scales larger than the average vortex spacing and the average vortex density varies smoothly as one goes from the vortex liquid at weak pinning to the vortex liquid at strong pinning.  They find no vortex glass, no vortex slush, and no multidomain glass of polycrystaline domains.  For the case of dilute well localized pins, they find again a vortex lattice melting transition with a similar shape as before (though quantitatively at a very different location, comparing amorphous to dilute pins with equal second moments of the random pinning potential).  Their vortex liquid state, however, now shows a clear polycrystaline structure with noticeable short range translational order extending on lengths larger than the average vortex spacing.  However they again find no first order transition within their vortex liquid phase, such as might define a region of vortex slush or multidomain glass as distinct from the vortex liquid.  Using a percolation criterion to define a crossover to glassy behavior within the vortex liquid phase, they find a line that is qualitatively similar in location to our vortex glass transition, as shown in Fig.~\ref{f1}.

\section*{Acknowledgments}

This work was supported by U.S. Department of Energy grant No. DE-FG02-06ER46298, by Swedish Research Council Contract No. 2007-5234, and by the resources of the Swedish High Performance Computing Center North (HPC2N).  We thank A. E. Koshelev for helpful discussion.

\section*{Appendix A}

We give a simplified summary of the Imry-Wortis scenario, as applied to our model, as follows.  Consider $T_{\rm m}(p)$ the nominal melting temperature of a system with disorder strength $p$.  Let 
\begin{equation}
\Delta f(p,T)\equiv f_{\rm lattice}(p,T)-f_{\rm liquid}(p,T)
\label{eADf}
\end{equation}
be the difference in free energy density between the vortex lattice and liquid states.  Consider now a volume $v$ in which, due to the random distribution of pins, the effective disorder strength $p_{\rm eff}$ is either greater than, or less than, the average $p$, $p_{\rm eff}=p\pm\Delta p$ ($\Delta p>0$).  Since $dT_{\rm m}/dp<0$, if $p_{\rm eff}>p$, the domain would lower its bulk free energy by {\it disordering} for some range of temperatures $\Delta T$ {\it below} $T_{\rm m}(p)$.  Similarly, if $p_{\rm eff}<p$, the domain would lower its bulk free energy by {\it ordering} for some range of $\Delta T$ {\it above} $T_{\rm m}$.  This is sketched in Fig.~\ref{f12} below.  
\begin{figure}[tbp]
\epsfxsize=8.6truecm
\epsfbox{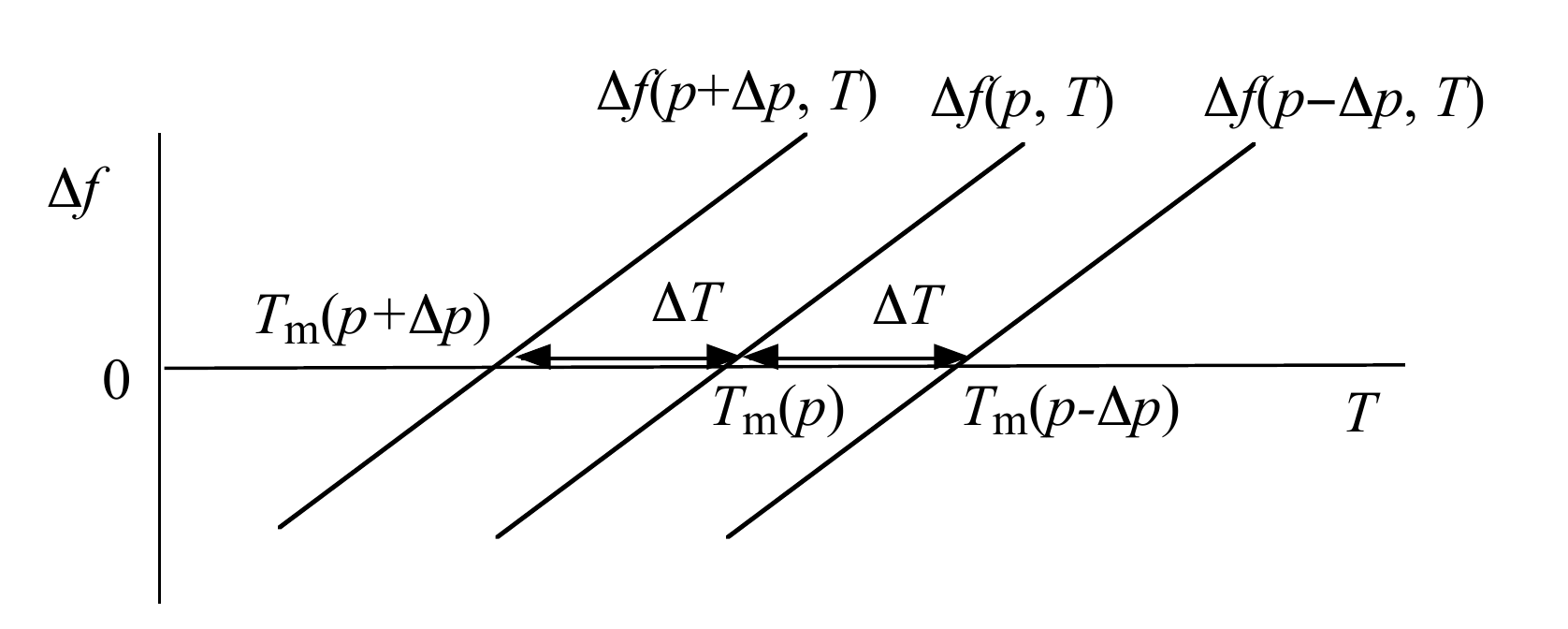}
\caption{
}
\label{f12}
\end{figure}
For the system siting at $T_{\rm m}(p)$, the {\it decrease} in the bulk free energy of such a domain would be,
\begin{eqnarray}
\Delta F_{\rm bulk}&=&\pm v\Delta f(p\pm\Delta p, T_{\rm m}(p))\nonumber\\
&\approx& v{\partial \Delta f(p,T_{\rm m}(p))\over\partial p}\Delta p\nonumber\\
&=&v\Delta Q(p,T_{\rm m}(p))\Delta p
\enspace.
\label{eADFb}
\end{eqnarray}
where the last line follows from our definition of $\Delta Q$ in Eq.~(\ref{DEQ}).

\begin{figure}[h]
\epsfxsize=6.0truecm
\epsfbox{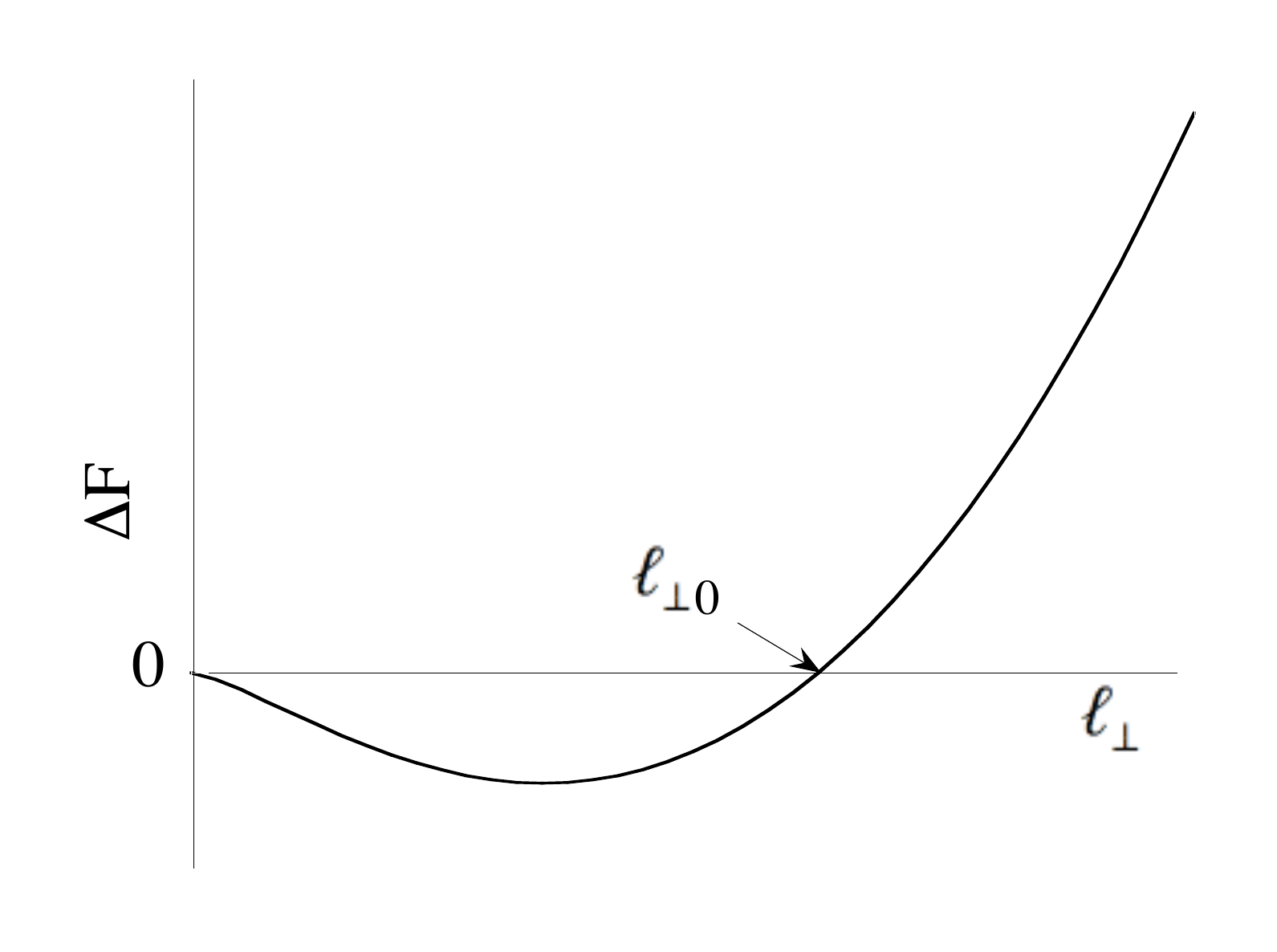}
\caption{
}
\label{f13}
\end{figure}

On the other hand, the formation such a domain would lead to an {\it increase} in surface free energy due to the resulting domain wall,
\begin{equation}
\Delta F_{\rm surface}\sim 2\sigma_z \ell_\perp^2 + 4\sigma_\perp\ell_\perp\ell_z\enspace,
\label{eADFs}
\end{equation}
where we take account of the anisotropy of the system by denoting $\ell_z$ and $\ell_\perp$ as the lengths of the domain parallel and perpendicular to the applied magnetic field, with $\sigma_z$ and $\sigma_\perp$ the surface tension between vortex lattice and liquid for surfaces with normal in these respective directions.  The domain will be unstable to flipping only if the total change in free energy is negative,
\begin{equation}
\Delta F = -\Delta F_{\rm bulk}+\Delta F_{\rm surface}<0\enspace.
\label{eADFtot}
\end{equation}
For simplicity, we will assume that the domains which form are such that the surface tension is equally distributed over all surfaces, so that,
\begin{equation}
\sigma_z \ell_\perp^2\sim\sigma_\perp \ell_\perp \ell_z,\quad {\rm so}\quad \ell_z\sim {\sigma_z\over\sigma_\perp}\ell_\perp\enspace,
\end{equation}
and so the volume of the domain is $v\sim \ell_\perp^2\ell_z=(\sigma_z/\sigma_\perp)\ell_\perp^3$.  The instability condition Eq.~(\ref{eADFtot}) then becomes,
\begin{equation}
\Delta F(\ell_\perp)=-{\sigma_z\over\sigma_\perp}\ell_\perp^3\Delta Q\Delta p + 6\sigma_z\ell_\perp^2<0\enspace.
\label{eADFtot2}
\end{equation}
Next, we can write that the typical variation in disorder $\Delta p$ for a domain of size $\ell_\perp^2\ell_z$ can be written as,
\begin{equation}
\Delta p = {\Delta T_{\rm m}\over |dT_{\rm m}/dp|}
={\Delta T_{\rm m}\Delta E\over T_{\rm m}\Delta Q}\enspace,
\label{eADp}
\end{equation}
where $\Delta T_{\rm m}$ is the variation in melting temperatures for domains of size $\ell_\perp^2\ell_z$ sampled from a system with average disorder $p$, and the last equality follows from the Clausius-Clapeyron relation, Eq.~(\ref{eCC2}).  If the system is self averaging, as is suggested by our results in Table~\ref{tab1}, then we expect
\begin{equation}
\Delta T_{\rm m}={\alpha\over \sqrt{\ell_\perp^2\ell_z}}={\alpha\over \sqrt{(\sigma_z/\sigma_\perp)}\ell_\perp^{3/2}}
\label{eADTm}
\end{equation}
where $\alpha$ is some constant.  Substituting Eqs.~(\ref{eADTm}) and (\ref{eADp}) into (\ref{eADFtot2}) then gives
\begin{equation}
\Delta F(\ell_\perp) = -\sqrt{\sigma_z\over\sigma_\perp}{\alpha\Delta E\over T_{\rm m}}\ell_\perp^{3/2}+6\sigma_z\ell_\perp^2<0\enspace.
\label{eADFtot3}
\end{equation}
We sketch $\Delta F(\ell_\perp)$ in Fig.~\ref{f13}.  Defining $\ell_{\perp 0}$ by $\Delta F(\ell_{\perp 0})=0$, we see that the system can be unstable only to the flipping of domains with transverse length $\ell_\perp<\ell_{\perp 0}$.  From Eq.~(\ref{eADFtot3}) we have,
\begin{equation}
\ell_{\perp 0} = {1\over\sigma_z\sigma_\perp}\left[{\alpha\Delta E\over 6 T_{\rm m}}\right]^2\enspace.
\label{eAellperp0}
\end{equation}
However, the above arguments only apply to domains which are well defined as such, i.e. they are at least as big as the correlation length that defines the minimum size of a domain.  Hence the instability condition for domain flipping becomes,
\begin{equation}
\xi_\perp<\ell_\perp<\ell_{\perp 0}\enspace,
\label{eAinst}
\end{equation}
and so, in particular, the system is {\it stable} against the flipping of local domains if $\xi_\perp>\ell_{\perp 0}$.   As the disorder is increased, one in general expects $\ell_{\perp 0}$ to increase.  So if at low disorder the system is stable, $\xi_\perp>\ell_{\perp 0}$, as the disorder increases one will eventually reach the condition $\xi_\perp=\ell_{\perp 0}$  and the system will first become unstable to domains on the size of the correlation length $\xi_\perp$.  As the disorder increases further, larger domains of size $\ell_\perp$, with $\ell_{\perp 0}>\ell_\perp>\xi_\perp$, will go unstable.

To test the Imry-Wortis scenario for our vortex line system, we therefore wish to compute the length $\ell_{\perp 0}$ of Eq.~(\ref{eAellperp0}).  We have already computed $T_{\rm m}$ and $\Delta E$, as shown in Figs.~\ref{f1} and \ref{f5} respectively.  We use our results from the $20\times 20\times 6$ size system, averaging $\Delta E/T_{\rm m}$ over the different realizations of randomness, and computing $\alpha = \Delta T_{\rm m}\sqrt{V}$ from the observed spread in melting temperatures, such as shown in Table~\ref{tab1} for the specific case of $p=0.12$.  It remains, therefore, to compute the surface tensions $\sigma_z$ and $\sigma_\perp$.

\begin{figure}[h]
\epsfxsize=6.0truecm
\epsfbox{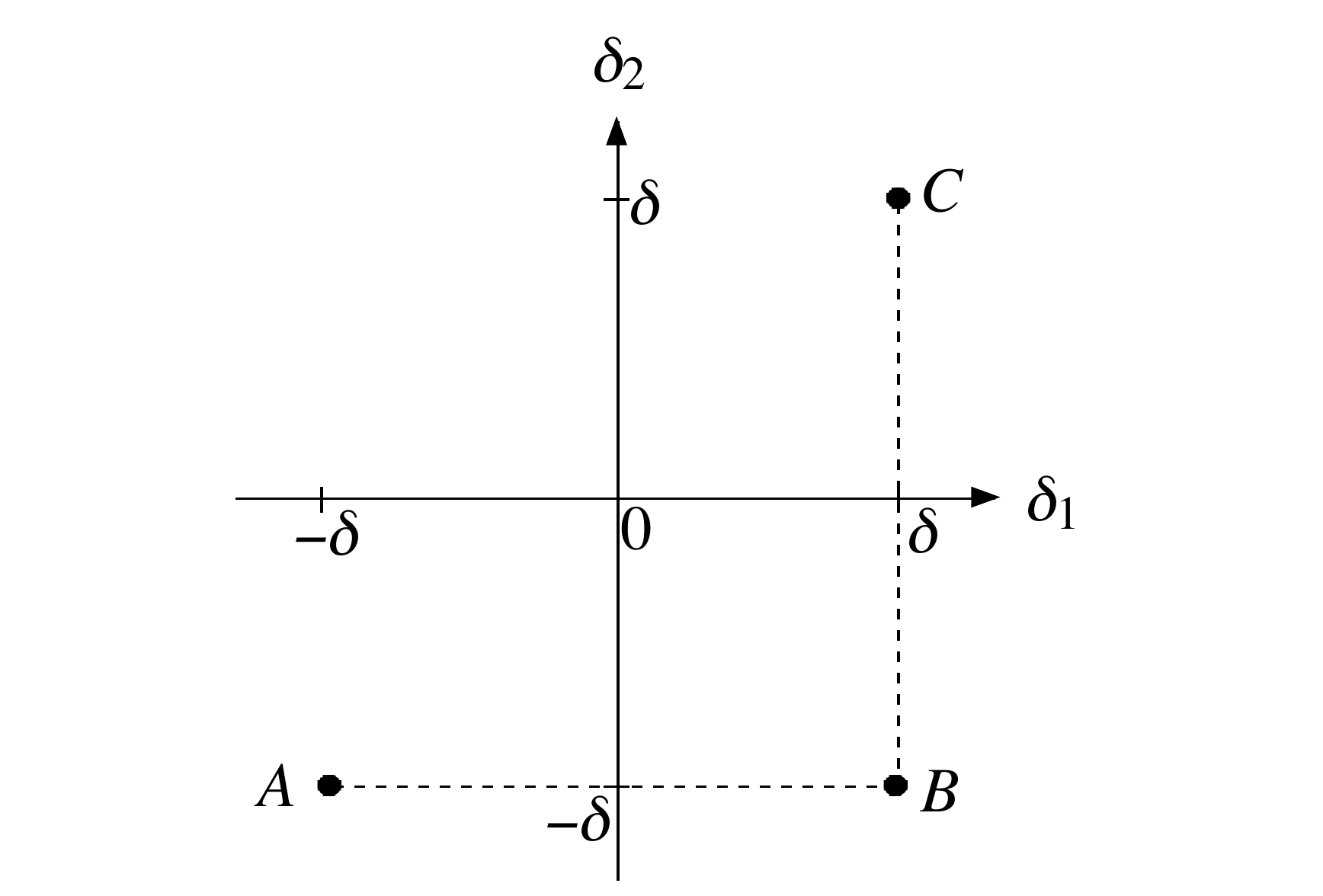}
\caption{
}
\label{f14}
\end{figure}

\begin{figure}[h]
\epsfxsize=8.0truecm
\epsfbox{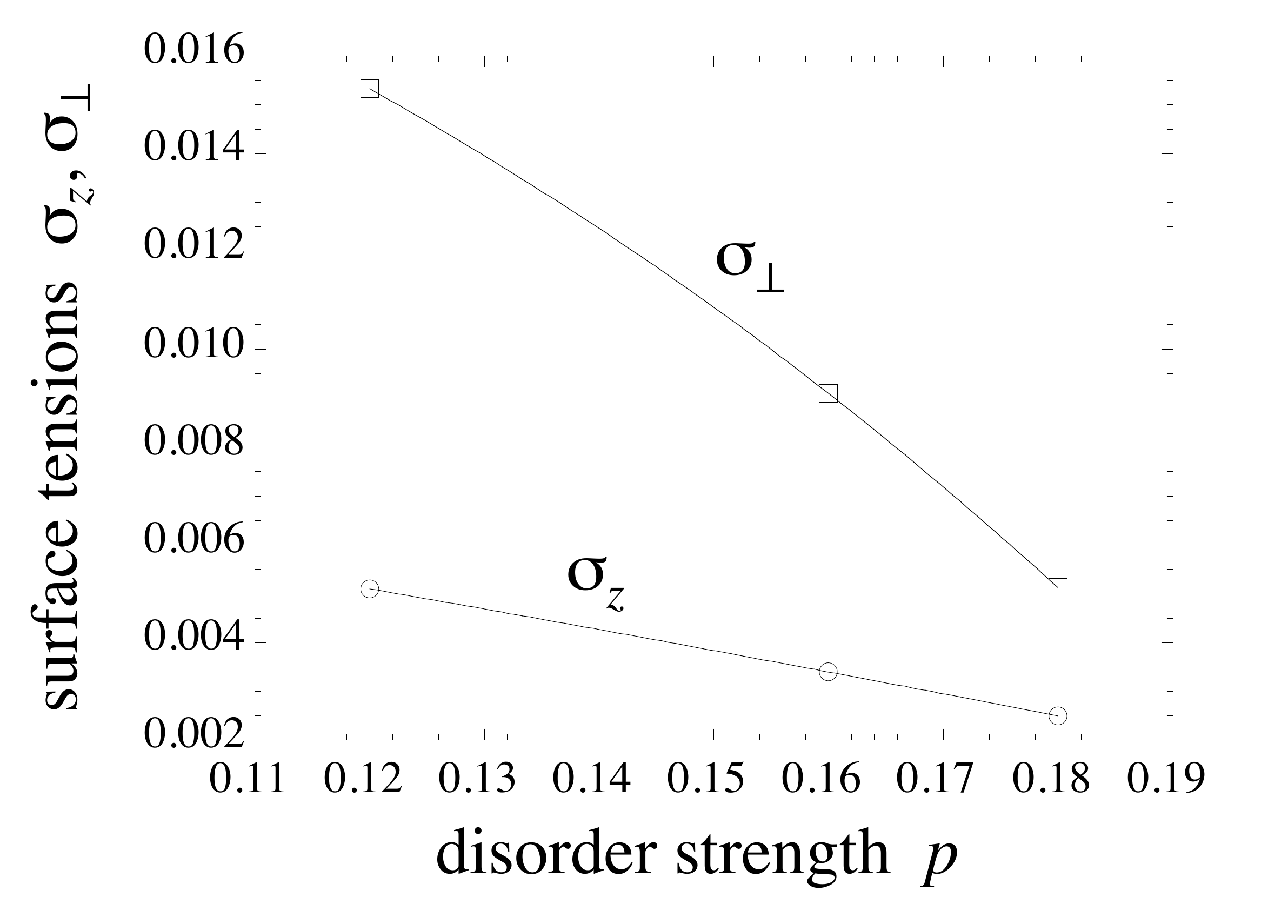}
\caption{Surface tensions between vortex lattice and vortex liquid states, for surfaces with normal parallel ($\sigma_z$) and perpendicular ($\sigma_\perp$) to the applied magnetic field, vs average disorder strength $p$.
}
\label{f15}
\end{figure}

To compute the surface tension we use a method based on the approach of Potvin and Rebbi.\cite{Potvin}  We take a given realization of the randomness for which we have previously determined the melting temperature $T_{\rm m}$.  We then take an exact copy of this system and join it to the original along the surface whose surface tension we seek to compute.  On one side, denoted as ``side 1",  we use couplings $J_{\perp 1}=J_\perp(1+\delta_1)$ and $J_{z1}=J_z(1+\delta_1)$ while on the other side, denoted as ``side 2", we use couplings $J_{\perp 2}=J_\perp(1+\delta_2)$ and $J_{z2}=J_z(1+\delta_2)$.  In this way we expect that exactly at $T_{\rm m}$ (as determined in the original system with $\delta_{1,2}=0$) if $\delta_{1,2}>0$, that side will be ordered, while if $\delta_{1,2}<0$, that side will be disordered.
Choosing $\delta_1=-\delta$ and $\delta_2=+\delta$ will thus create an interface between ordered and disordered halves of the total system.  Consider now a trajectory in the $(\delta_1,\delta_2)$ plane, as shown in Fig.~\ref{f14}.  
In this figure, point $A$ is a totally disordered system, point $C$ is a totally ordered system, and point $B$ has side 1 ordered and side 2 disordered.  The total free energy of the system at point $B$ can be written as
\begin{equation}
F_B={1\over 2}F_A + {1\over 2}F_C+2\Sigma
\label{eA1}
\end{equation}
where $F_A$ and $F_C$ are the total free energies at points $A$ and $C$, and $\Sigma$ is the total surface free energy of one interface between the ordered and disordered halves.  The factor $2\Sigma$ appears since our periodic boundary conditions necessarily creates two interfaces equally spaced by half the length of the total system.   From this we have,
\begin{equation}
4\Sigma=[F_B-F_A]+[F_B-F_C]\enspace.
\label{eA2}
\end{equation}
The surface tension between coexisting disordered and ordered phases at the same transition temperature $T_{\rm m}$ is then obtained from $\Sigma$, taking in principle the limit of $\delta\to 0$. 
To evaluate the free energy differences in the above equation we use,
\begin{equation}
F_B-F_A=\int_{-\delta}^{+\delta}d\delta_1 {\partial F(\delta_1,\delta_2)\over\partial \delta_1}
=\int_{-\delta}^{+\delta}d\delta_1 {E_1(\delta_1,\delta_2)\over 1+\delta_1}\enspace,
\label{eA3}
\end{equation}
where $E_1(\delta_1,\delta_2)$ is the total energy of side 1 at the specified couplings.  A similar expression can be derived for $F_B-F_C$.  Simulating at points along the trajectory $A\to B\to C$ we then integrate the energies $E_1$ and $E_2$ to compute the surface tension,
\begin{equation}
\sigma={\Sigma\over A}={1\over 4A}\left[\int_{-\delta}^{+\delta}d\delta_1{E_1(\delta_1,\delta_2)\over 1+\delta_1} - \int_{-\delta}^{+\delta}d\delta_2{E_2(\delta_1,\delta_2)\over 1+\delta_2}\right]\enspace,
\label{eA4}
\end{equation}
where $A$ is the total area of one interface.   We implement this procedure on a $20\times20\times 6$ system doubled in the $z$ direction (to make a $20\times20\times 12$ system) so as to compute $\sigma_z$, and doubled in the $x$ direction (to make a $40\times20\times 6$ system) so as to compute $\sigma_\perp$.  We use a value $\delta=0.1$ in order to get reasonable results.  Our results are averaged over 8 independent realizations of the random disorder (only 7 for $p=0.18$).  We plot our results in Fig.~\ref{f15}.
As expected, $\sigma_z$ and $\sigma_\perp$ decrease as the disorder strength $p$ increases.  For our parameters of anisotropy and vortex line density we find $\sigma_z\approx \sigma_\perp/3$.

We summarize the pieces of our calculation of $\ell_{\perp 0}$ in Table~\ref{tab2}.  The values for $\sigma_z$ and $\sigma_\perp$ are obtained as described above.  Values for $[\Delta E/T_{\rm m}]$ are obtained averaging over  careful equilibrations of 20, 8, and 7 different realizations of the random disorder for $p=0.12$, 0.16 and 0.18 respectively, for a $20\times20\times 6$ system.  Because the spread in melting temperatures $\Delta T_{\rm m}$ is the quantity that is most sensitive to the fact that we sample only over a rather small number of random realizations, for $p=0.16$ ($0.18$) we have tried to do better than the 8 (7) realizations we have carefully equilibrated by computing $\Delta T_{\rm m}$ from 16 random realizations where we determine $T_{\rm m}$ from shorter runs and more qualitative methods.  We then use $\alpha=\Delta T_{\rm m}\sqrt{20^2\times 6}$.
\begin{table}[htdp]
\caption{Values that enter our calculation of $\ell_{\perp 0}$ from Eq.~(\ref{eAellperp0}). } 
\begin{center}
\begin{tabular}{|c|c|c|c|c|c|}
\hline
$p$ & $\Delta T_{\rm m}$ & $[\Delta E/T_{\rm m}]$ & $\sigma_z$ & $\sigma_\perp$ & $\ell_{\perp 0}$ \\
\hline
0.12 & 0.0037 & 0.124 & 0.0051 & 0.0153 & 0.18\\
0.16 & 0.0060 & 0.074 & 0.0034 & 0.0091 & 0.42\\
0.18 & 0.0090 & 0.076 & 0.0025 & 0.0051 & 2.41\\
\hline
\end{tabular}
\end{center}
\label{tab2}
\end{table}
%

Our results show $\ell_{\perp 0}$ to be an increasing function of disorder strength $p$, as expected.  However, for the system to become unstable to the flipping of domains it is necessary that $\ell_{\perp 0}>\xi_\perp$.  For our vortex density of $f=1/5$ we estimate that $\xi_\perp$ at $T_{\rm m}$ is at least as large as the average vortex spacing, $a_{\rm v}=1/\sqrt{5}\simeq 2.2$. This seems consistent with the real space images of Fig.~\ref{f10} were we see ordered regions of at least this size in the liquid, and disordered regions of at least this size in the lattice. Thus we have $\ell_{\perp 0}\gtrsim\xi_\perp$ only for our strongest disorder strength $p=0.18$, where our results are perhaps the least accurate.  In contrast, our phase diagram of Fig.~\ref{f1} shows that the mixed state is already observed for disorder strengths  as low as $p=0.14$.  

It should be noted that the above analysis is based only on typical ``root mean square" behaviors.  Correlations between bulk and surface free energies of domains may enhance the effect over what we have estimated above.  For example, a domain of lattice may flip to the liquid state in a region where the vortex pinning is locally stronger than on average; but in such a region, Fig.~\ref{f15} shows that the surface tension is lower than average, thus reducing the energy cost of such a flip from that considered in our arguments above.  Domains may also flip in regions of the system where the value of the local disorder strength lies further out in the tails of the disorder strength distribution, rather than near the root mean square value.  This might explain why we see our intermediate mixed states more easily when we increase the system size, thus affording a wider sampling of the disorder strength distribution within any given  single sample.  
The results of our Imry-Wortis analysis thus show the right trends for explaining our intermediate mixed states, however in the absence of a clear quantitative agreement, $\ell_{\perp 0}\sim\xi_\perp$, we must regard our results as still inconclusive.

\end{document}